\documentclass[acmlarge, screen, nonacm, authorversion]{acmart}
\usepackage{multirow}
\usepackage{subcaption}
\usepackage{booktabs}
\usepackage{wrapfig,graphics,graphicx}
\usepackage{amsfonts}
\usepackage{amsmath}
\usepackage{url}
\DeclareMathOperator*{\argmax}{argmax}

\AtBeginDocument{%
  \providecommand\BibTeX{{%
    \normalfont B\kern-0.5em{\scshape i\kern-0.25em b}\kern-0.8em\TeX}}}

\begin{document}

\title{Plug-and-Play Multilingual Few-shot Spoken Words Recognition} 

\author{Aaqib Saeed}
\email{a.saeed@tue.nl}
\affiliation{%
  \institution{Eindhoven University of Technology}
  \city{Eindhoven}
  \country{The Netherlands}}

\author{Vasileios Tsouvalas}
\affiliation{%
  \institution{Eindhoven University of Technology}
  \city{Eindhoven}
  \country{The Netherlands}}

\renewcommand{\shortauthors}{Saeed et al.}

\begin{abstract}
As technology advances and digital devices become prevalent, seamless human-machine communication is increasingly gaining significance. The growing adoption of mobile, wearable, and other Internet of Things (IoT) devices has changed how we interact with these smart devices, making accurate spoken words recognition a crucial component for effective interaction. However, building robust spoken words detection system that can handle novel keywords remains challenging, especially for low-resource languages with limited training data. Here, we propose PLiX, a multilingual and plug-and-play keyword spotting system that leverages few-shot learning to harness massive real-world data and enable the recognition of unseen spoken words at test-time. Our few-shot deep models are learned with millions of one-second audio clips across 20 languages, achieving state-of-the-art performance while being highly efficient. Extensive evaluations show that PLiX can generalize to novel spoken words given as few as just one support example and performs well on unseen languages out of the box. We release models and inference code to serve as a foundation for future research and voice-enabled user interface development for emerging devices.
\end{abstract}



\maketitle

\section{Introduction}\label{sec:introduction}

The proliferation of mobile, wearable, and other IoT devices has changed the way we communicate, navigate, and carry out various tasks, making them an integral part of our daily lives. The adoption of these emerging devices in our routine demands seamless communication between humans and machines. Consequently, the accurate recognition of spoken words\footnote{We use words, terms and classes interchangeably.}, or keyword spotting (KWS), has become a fundamental aspect of interaction between individuals and such devices as it facilitates a natural dialogue. Specifically, it is an essential component of virtual assistants, such as Siri and Alexa, which have revolutionized the way users interact with their devices, making the experience more natural, personalized and intuitive. Keyword spotters are relatively low-resource wake-word engines that constantly monitor audio, listening for specific trigger phrases in the audio stream. Once the phrase is detected, the system wakes up to activate a service or perform a task, such as turning on the TV or initiate an engagement with an in-car system while driving. 

With the advancement of deep learning algorithms and speech recognition technologies, keyword spotting has become widespread, enabling improved interaction between users and devices. This technology is increasingly used in a wide range of applications, including smart homes, automotive systems, and healthcare. However, building robust and accurate models for multiple languages remains a challenging task, particularly for low-resource ones that lack large amounts of labeled data for training. Despite significant advances in deep neural network-based spoken word recognition, traditional learning methods require large amounts of data for training, which can be time-consuming, expensive, and entail retraining of models to accommodate novel keywords depending on evolving user needs. Likewise, traditional learning methods may require significant amounts of computational resources, making them unsuitable for resource-constrained devices, and on-device personalization is not feasible either. Thus, there is a need for developing more sophisticated and effective keyword spotting systems to cater to a multilingual world and rapid adaptation to handle novel words at inference-time. The development of such a spoken word recognition system can play a vital role in shaping the future of human-technology interaction, enabling seamless and natural communication between individuals and digital devices across different languages. 

\begin{figure}[!htbp]
    \centering
    \includegraphics[width=\textwidth]{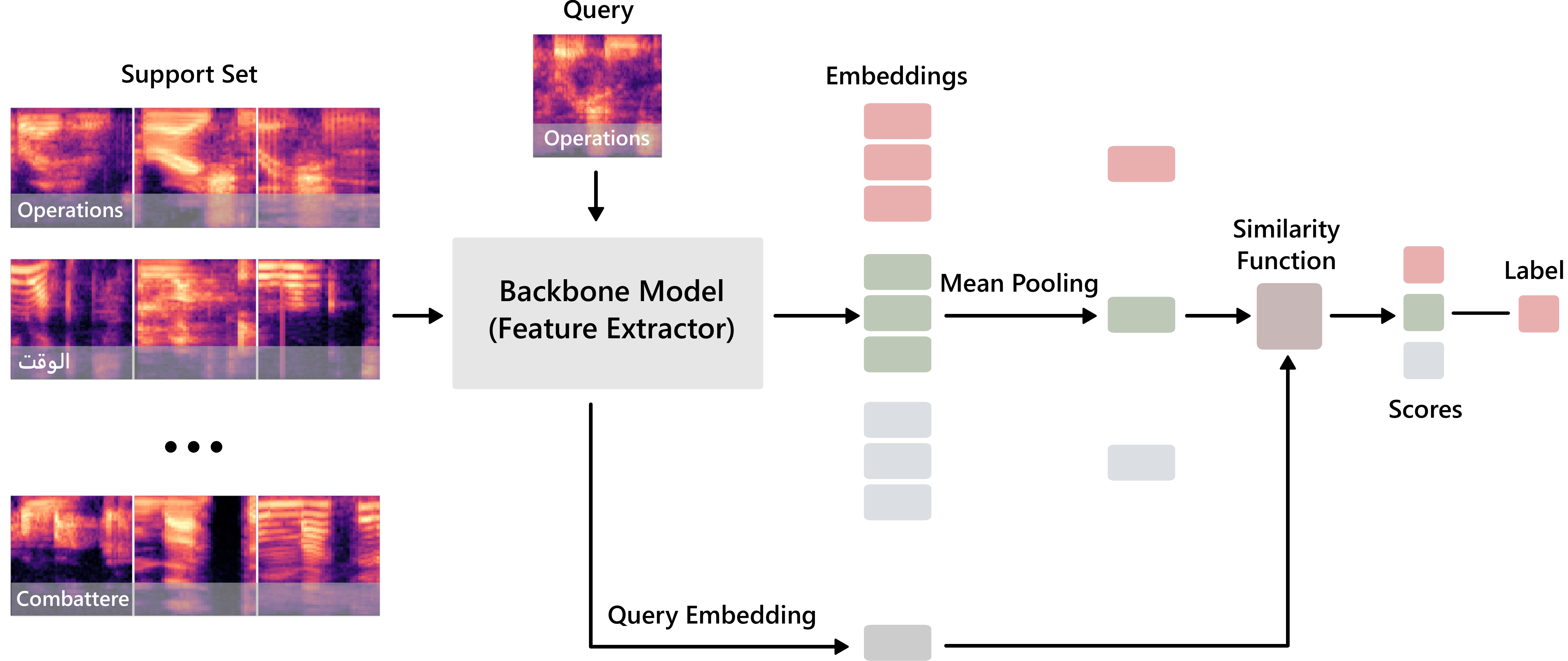}
    \caption{Illustration of few-shot training paradigm with Prototypical Networks~\cite{snell2017prototypical} for multilingual spoken words recognition.}
    \label{fig:main_overview}
\end{figure}

Recent studies have posited methods capable of learning deep models that generalize to unseen tasks with limited training data, using an approach termed \textit{few-shot learning}. In this realm, a classifier must learn to identify novel classes given only a small number of exemplars from each class. Conventional few-shot learning techniques frame the problem as a closed-set classification task, where the objective is to label a query sample as one of $C$ known classes given $K$ exemplars per class. It has shown great success in solving meta-learning problems ranging from sound event detection to~\cite{wang2020few,shi2020few} image classification~\cite{snell2017prototypical,sung2018learning,santoro2016meta,ravi2017optimization}. The key idea is to learn a model that can quickly adapt to new tasks by learning from just a few examples, which enables the approach to tackle a variety of tasks across different domains. Here, to address the lack of a general-purpose, language-agnostic, and easy-to-use keyword spotter, we propose \textbf{PLiX} (see Figure~\ref{fig:main_overview} for illustration), a system for leveraging few-shot learning to harness massive real-world data and enable the recognition of unseen spoken words. Overall, our work makes the following main contributions:

\begin{itemize}
    \item We develop PLiX, a \textit{general-purpose}, \textit{multilingual} , and \textit{plug-and-play}, few-shot keyword spotting system trained and evaluated with more than $12$ million one-second audio clips sampled at $16$kHz.
    \item Leverage state-of-the-art neural architectures to learn few-shot models that are high performant while being efficient with fewer learnable parameters. 
    \item A wide-ranging set of evaluations to systematically quantify the efficacy of our system across $20$ languages and thousands of classes (i.e., words or terms); showcasing generalization to unseen words at test-time given as few as one support example per class. 
    \item We demonstrate that our model generalizes exceptionally well in a one-shot setting on $5$ unseen languages. Further, in a cross-task transfer evaluation on a challenging FLEURS benchmark~\cite{fleurs2022arxiv}, our model performs well for language identification without any retraining. 
    \item To serve as a building block for future research on spoken word detection with meta-learning and enable product development, we release model weights and inference code at: \url{https://github.com/FewshotML/plix}.
\end{itemize}

\section{Methodology}
\label{sec:methodology}

\subsection{Preliminaries}

\subsubsection{Few-shot Learning}
Few-shot learning (FSL) is a subfield of machine learning that deals with the problem of learning to perform tasks (such as object classification) when only a limited amount of labeled data is available. In classic learning regime, the primary objective is to acquire a model that can generalize well to previously unseen data of the same classes by leveraging a substantial amount of labeled training data. However, in numerous real-world applications, obtaining extensive labeled data is often an arduous, expensive, and time-consuming process, which renders FSL an attractive paradigm for developing data-efficient models. The appeal of few-shot learners lies in their ability to rapidly adapt to new tasks with minimal labeled data, thereby providing an attractive and practical alternative to standard learning methods. Recent advancements in FSL have been attained by the adoption of an episodic paradigm. Specifically, consider a scenario in which we possess a sizable labeled dataset for a particular set of classes $C_{train}$. However, our ultimate objective is to create classifiers for a distinct set of new classes $C_{test}$, for which only a limited number of labeled examples are available. In this regard, the episodic paradigm provides a framework for emulating the types of few-shot problems that are likely to be encountered during testing while simultaneously harnessing the abundance of labeled data that is accessible for the classes belonging to $C_{train}$.

The training of models in the FSL paradigm is performed on $K$-shot, $N$-way episodes, which involves first sampling a small subset of N classes from the set of available classes $C_{train}$. Subsequently, a training (support) set $S$ and a test (query) set $Q$ are generated. The support set $S$ encompasses K examples from each of the N classes, i.e., $S=\{(x_1,y_1), \ldots, (x_{N\times K},y_{N\times K})\}$, where $x_i\in\mathbb{R}^D$ denotes an input vector with dimensionality D and $y_i\in{1,\ldots, N}$ is a class label. On the other hand, the test set $Q$ comprises diverse examples from the same N classes, i.e., $Q=\{(x^*_1,y^*_1), \ldots, (x^*_T,y^*_T)\}$, where $x^*_i\in\mathbb{R}^D$, $y^*_i\in\{1,2,\ldots, N\}$, and $T$ denotes the number of query examples. To train on such episodes, the model $f_\theta(\cdot)$ is presented with the support set $S$, and its parameters are updated to minimize a specified loss function $\min_{\theta} \sum_{E \in D} L(f_\theta(S), Q) $ that quantifies the discrepancy between the model's predictions and the ground truth labels for the examples in the query set $Q$. This training process involves iteratively adjusting the model's internal representations to improve its ability to generalize to new classes with limited labeled data at test time. Therefore, learning within this framework is often referred to as \textit{learning to learn} or \textit{meta-learning}~\cite{thrun1998lifelong}, as it involves training the model to learn how to quickly adapt to unseen tasks with limited labeled data, similar to how humans can learn novel tasks with few demonstrations by leveraging their prior knowledge and experience.

\subsubsection{Prototypical Network}
Prototypical Network (or ProtoNet) is a simple yet highly effective approach to few-shot learning, which was proposed by Snell et al.~\cite{snell2017prototypical}. The central concept behind ProtoNets is the computation of a prototype for each class in the support set. These prototypes serve as a representative summary of the class and are then used for the classification of instances in the query set. In essence, ProtoNet maps input samples to a metric (or embedding) space, where the distances between prototypes and query examples are used to perform classification. 

The learning procedure beings with the computing of the prototype for each class in the support set $S$. Let $S_i$ denote the set of instances in $S$ belonging to class $i$, and let $f_{\theta}$ denote a neural network with learnable parameters $\theta$. The prototype for class $i$ is then defined as the mean embedding of the instances in $S_i$: $\mathbf{p}_{i} = \frac{1}{|S_i|} \sum_{(x, y) \in S_i} f_{\theta}(x)$, where $|\cdot|$ denotes the cardinality of a set. Given the prototypes $\mathbf{p}_i$, the classification score for an instance $\mathbf{x}$ in the query set can be computed as the negative Euclidean distance between $\mathbf{x}$ and each prototype $\mathbf{p}_i$: $\mathbf{s}_{i} = -\left\lVert f_{\theta}(\mathbf{x})-       \mathbf{p}_i\right\rVert_2^2$, where $\left\lVert \cdot \right\rVert_2$ denotes the Euclidean norm. Finally, the predicted class for $\mathbf{x}$ is the one with the highest classification score: $\hat{y} = \argmax_{i \in \{1, \dots, N\}} s_i$.
ProtoNet can be trained end-to-end by minimizing a cross-entropy loss over the predicted class probabilities for the instances in the query set and performing a gradient descent update. Let $Q_i$ denote the set of instances in the query set belonging to class $i$, and let $y_j \in {1, \ldots, N}$ denote the true class label of instance $\mathbf{x}_j$ in $Q$ while $\mathbf{1}(\cdot)$ is the indicator function. Then, the loss for a single episode can be written as: $\mathcal{L}=-\sum_{j=1}^{\left|Q\right|}\sum_{i=1}^{N} 1(y_j=i) \log \frac{\exp(s_i)}{\sum_{k=1}^{N} \exp(s_k)}$.

ProtoNet learns to classify instances in the query set by leveraging a set of prototypes computed from the support set examples. After training, its generalization performance is assessed on test set episodes, which include instances from classes in $C_{test}$ instead of $C_{train}$. For each test episode, the Prototypical Network uses the predictor produced by the support set $S$ to classify each query input $\mathbf{x}^*$ into the most likely class $\hat{y} = \argmax_c p(c|\mathbf{x}^*, {\mathbf{p}_i})$. This approach has showcased state-of-the-art performance on few-shot learning benchmarks. Due to its simplicity and scalability, we leverage ProtoNet to solve the task of in-the-wild spoken words recognition.

\subsection{Few-shot Recognition System}
\subsubsection{Overview}
\begin{figure}[!htbp]
    \centering
    \includegraphics[width=0.85\textwidth]{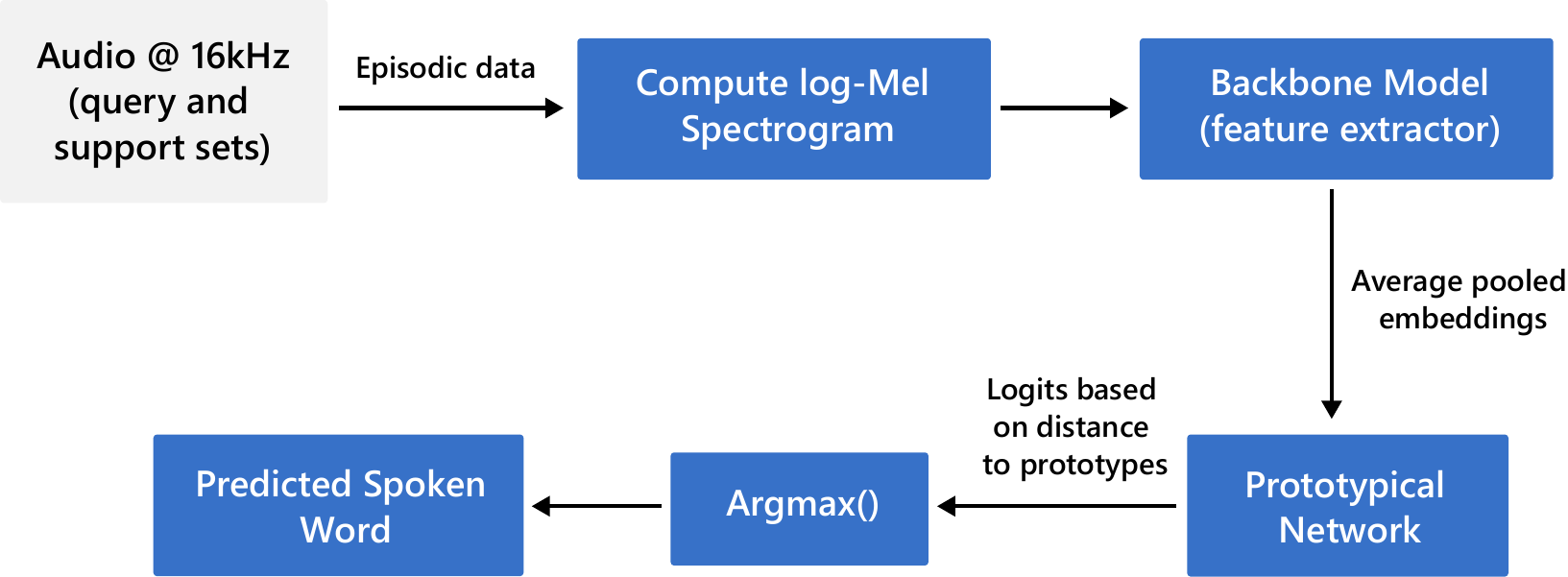}
    \caption{Overview of PLiX System Design.}
    \label{fig:sys_overview}
\end{figure}

We provide a high-level overview of the PLiX's system design in Figure~\ref{fig:sys_overview}. It comprises an acoustic model, a neural network that uses visual representation of sound to learn useful features from massive multilingual audio data (Section~\ref{subsec:ama}). Our few-shot learning system based on ProtoNet learns via episodic training of spoken words to learn to generalize to unseen words at test-time (Section~\ref{subsec:tp}). It uses an audio segments of one-second sampled at $16$kHz as query and support examples to detect spoken words of interest. The designed system can be leveraged in a variety of settings (Section~\ref{subsec:is}) to develop novel applications with voice as interface in a plug-and-play manner.    

\subsubsection{Acoustic Model Architecture}
\label{subsec:ama}
We leverage state-of-the-art neural network architectures as `encoders' or `feature extractors' to learn generalizable representations from audio. We focus on neural architectures that are effective both in terms of rapid training and inference (including on resource-constrained devices and cloud). Our few-shot learner use EfficientNet-v2~\cite{tan2021efficientnetv2} and TinyNet~\cite{han2020model} as encoders or backbone network in ProtoNets. We further refer the former as a `base' model and the latter as a `small' network based on their sizes. For model input, we use a visual representation of an audio signal, known as `log-Mel spectrogram'. We obtain it by computing the short-time Fourier transform of an audio signal, followed by filtering the resulting spectra through a bank of mel-frequency filters. The logarithm of the resulting mel-filterbank energies is then computed, resulting in a compact, yet informative representation of the spectral content of the audio signal over time. The log-Mel spectrogram has several advantages over other spectral representations, including their ability to capture both spectral and temporal features of the signal, as well as their robustness to noise and distortion. Specifically, it has been used in a wide range of applications, including speech recognition, music genre classification, and sound event detection. 

The EfficientNet-v2 is designed to achieve better performance with fewer parameters and faster training than its predecessor. It introduces several key improvements, including a scaling method called `compound scaling' to achieve better performance with smaller and faster models. Compound scaling combines the scaling of depth, width, and input resolution to find the optimal trade-off between accuracy and computational efficiency. Furthermore, it introduces a new type of convolutional block called `Fused-MBConv' that combines multiple types of convolutional layers into a single block. Fused-MBConv is more efficient than the original MBConv block used in EfficientNet as it eliminates redundant operations and reduces the number of parameters. There are different variants of the main architecture, mostly differing in number of parameters and utilized pretraining dataset. Due to its greater generalization ability, we use EfficientNet-v2 (Medium) with $52.8$ million parameters as our `base' model for learning multilingual and English-only few-shot models. 

To design even smaller yet highly performant few-shot learner, we employ TinyNets as our `small' model that we use to train spoken language specific models due to their tiny memory footprint. TinyNets are a family of neural networks designed to achieve high accuracy with a small number of parameters. These networks are particularly suitable for deployment on resource-constrained devices, such as smartphones or embedded systems, where memory and processing power are limited. The main novelty of TinyNets involves manipulating three key dimensions of a network: resolution, depth, and width. It is achieved through an approach called, the `Model Rubik's Cube'~\cite{han2020model}, where each dimension corresponds to a face of the cube that can be twisted to generate different network configurations with varying sizes and accuracies. In our work, we employ smallest model named, TinyNet-E with $761,396$ parameters as a `feature extractor' that the ProtoNet use as a backbone. 

\begin{figure}[t]
\subfloat{\includegraphics[width=0.33\textwidth]{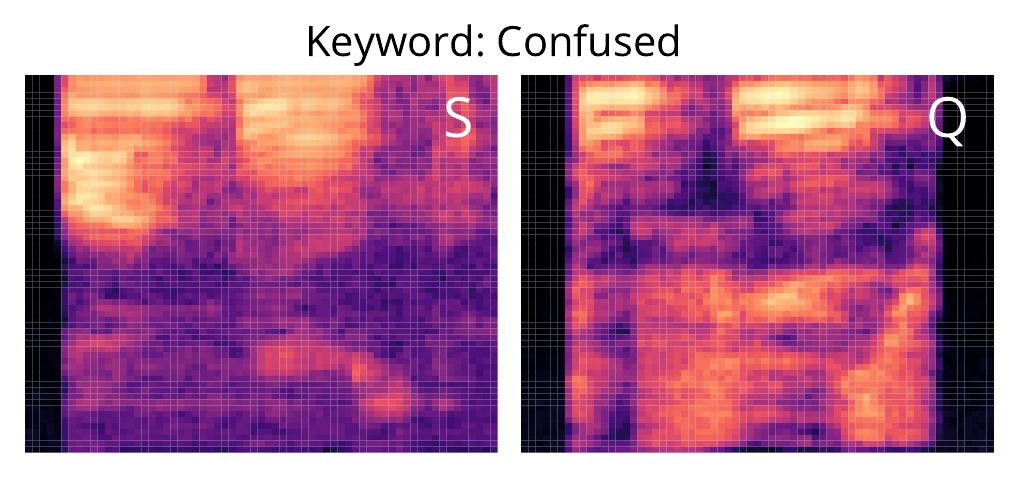}}  
\subfloat{\includegraphics[width=0.33\textwidth]{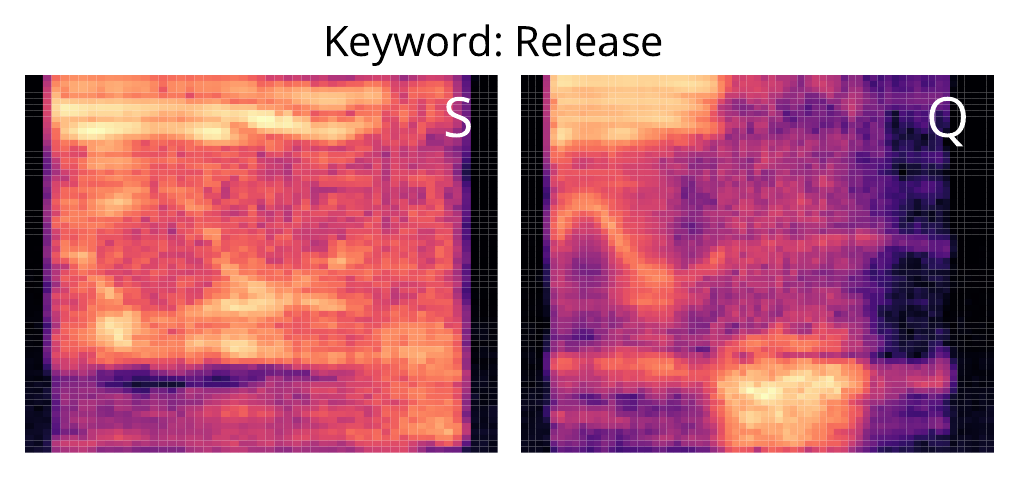}}
\subfloat{\includegraphics[width=0.33\textwidth]{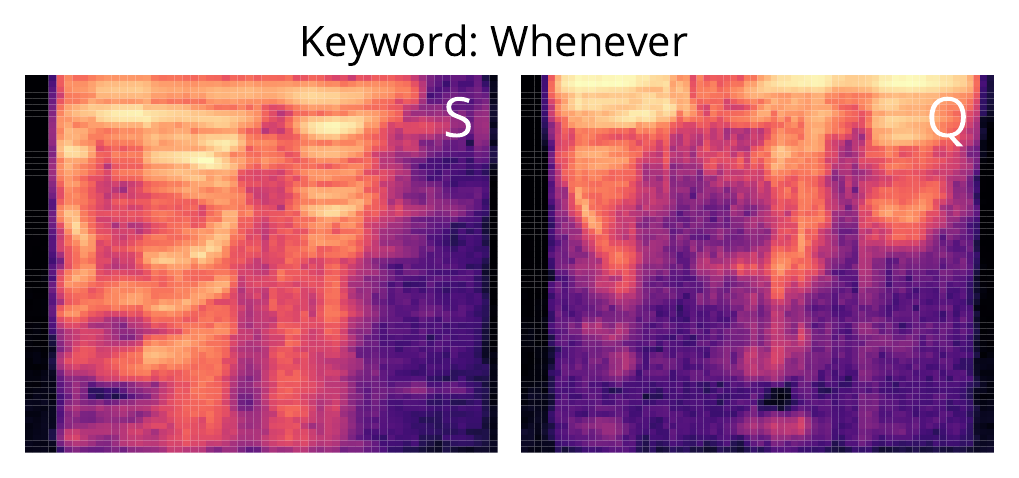}} \\

\subfloat{\includegraphics[width=0.33\textwidth]{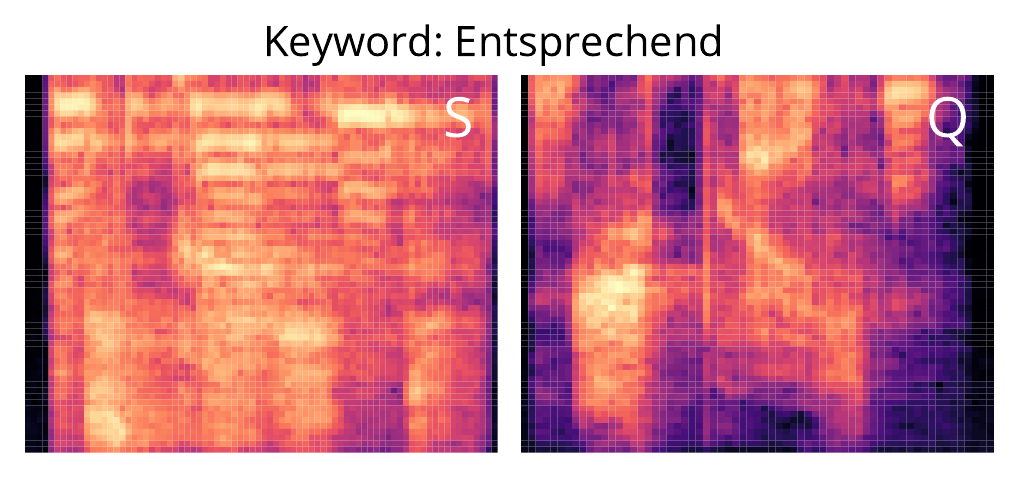}}  
\subfloat{\includegraphics[width=0.33\textwidth]{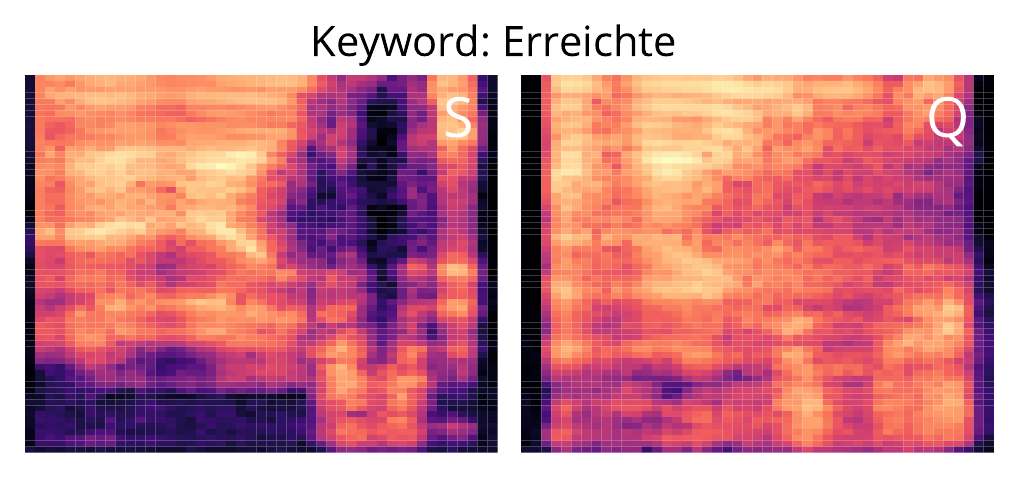}}
\subfloat{\includegraphics[width=0.33\textwidth]{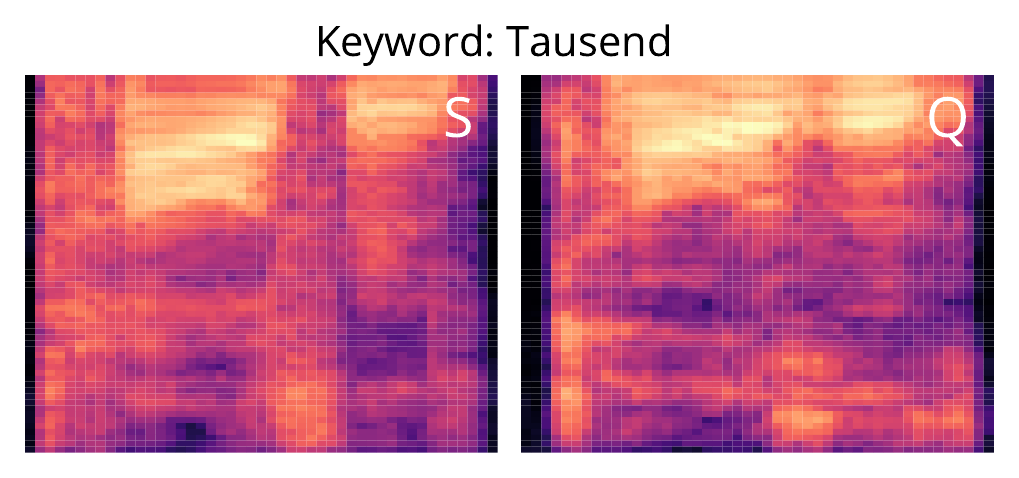}} \\

\subfloat{\includegraphics[width=0.33\textwidth]{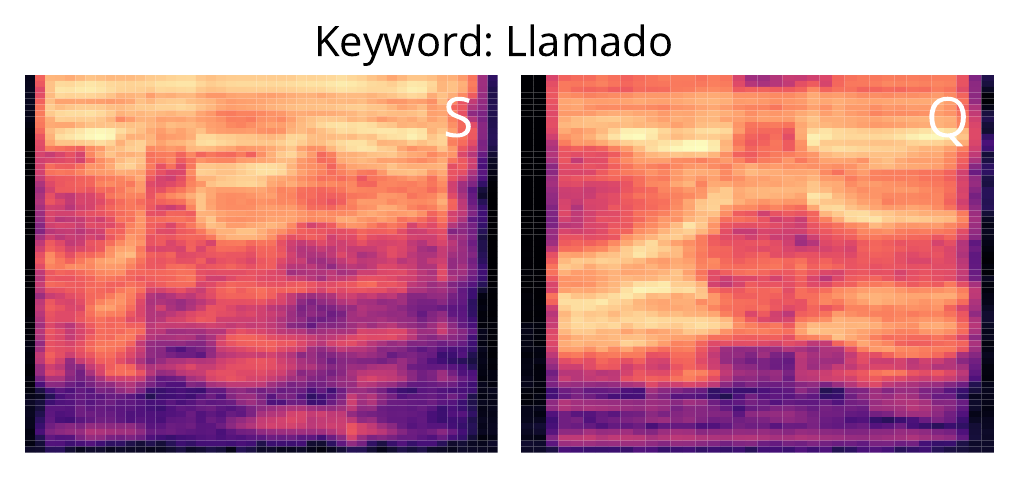}}  
\subfloat{\includegraphics[width=0.33\textwidth]{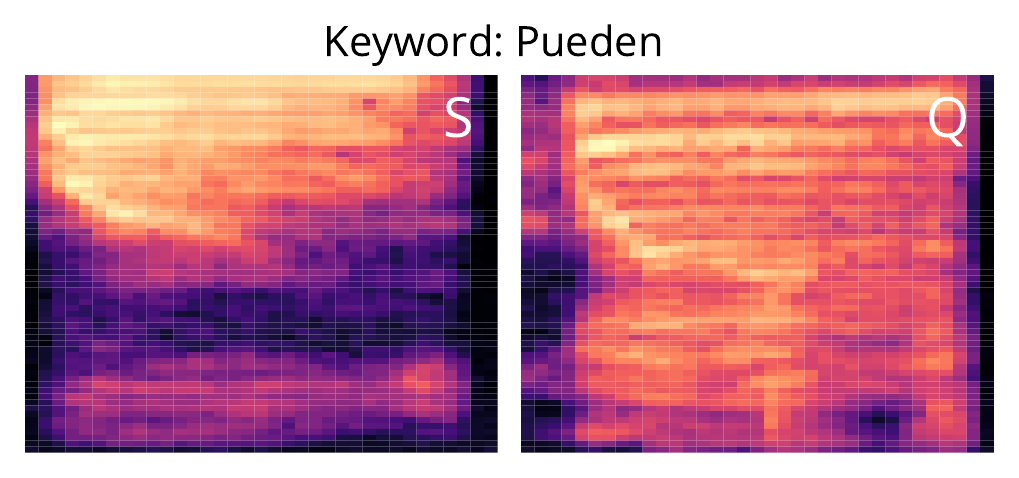}}
\subfloat{\includegraphics[width=0.33\textwidth]{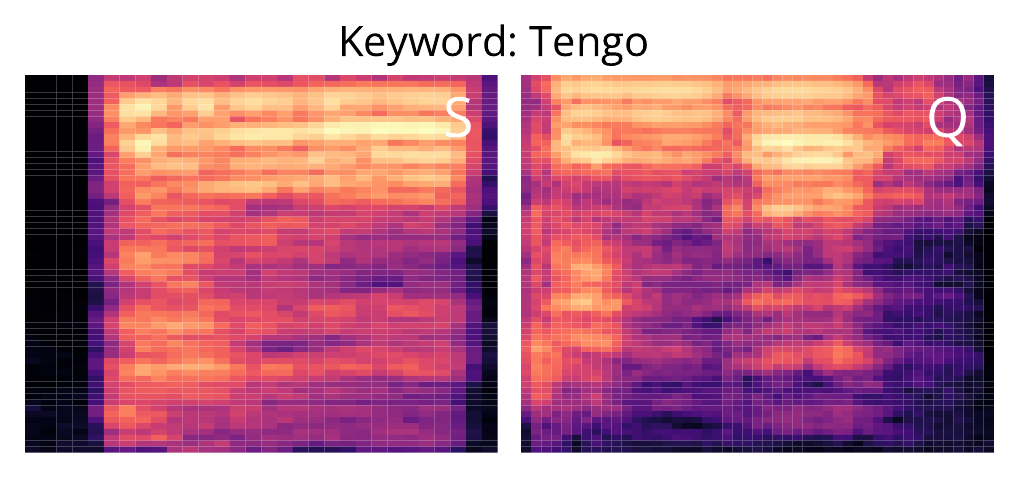}} \\

\subfloat{\includegraphics[width=0.33\textwidth]{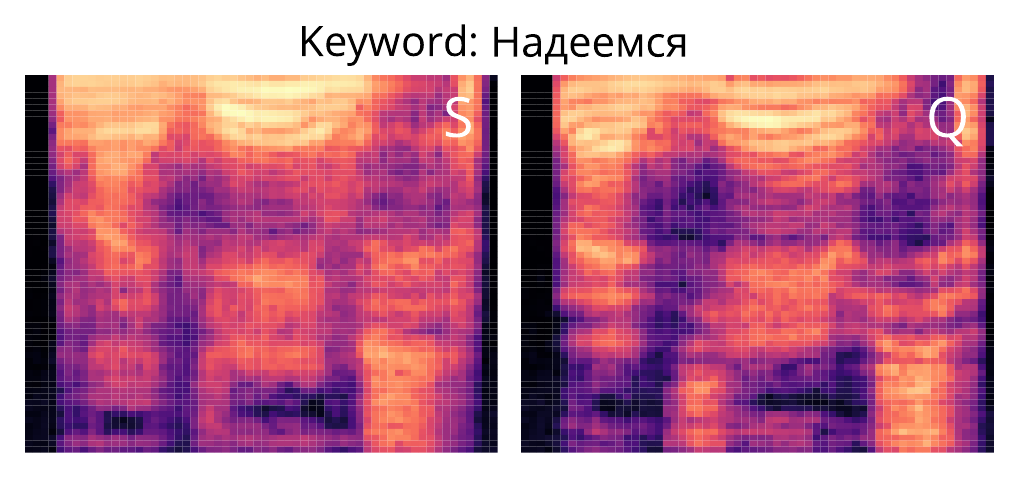}}  
\subfloat{\includegraphics[width=0.33\textwidth]{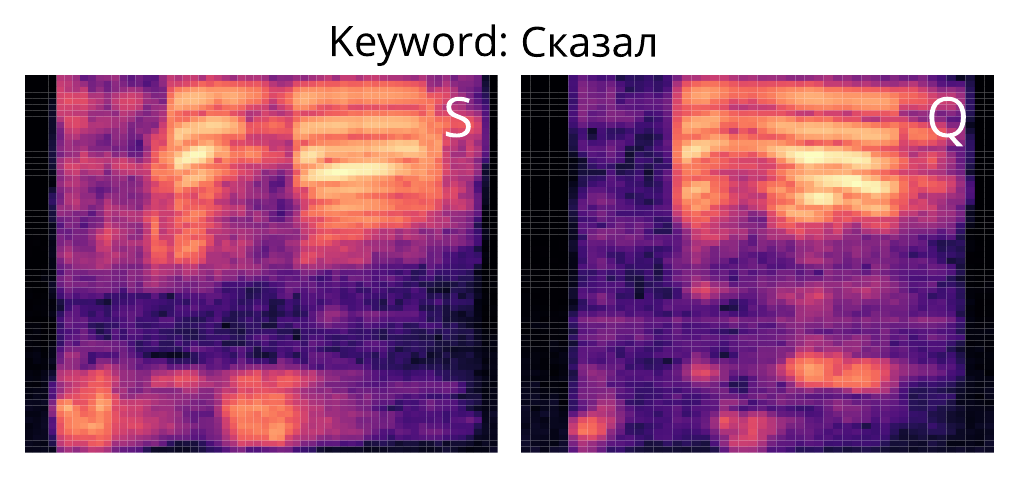}}
\subfloat{\includegraphics[width=0.33\textwidth]{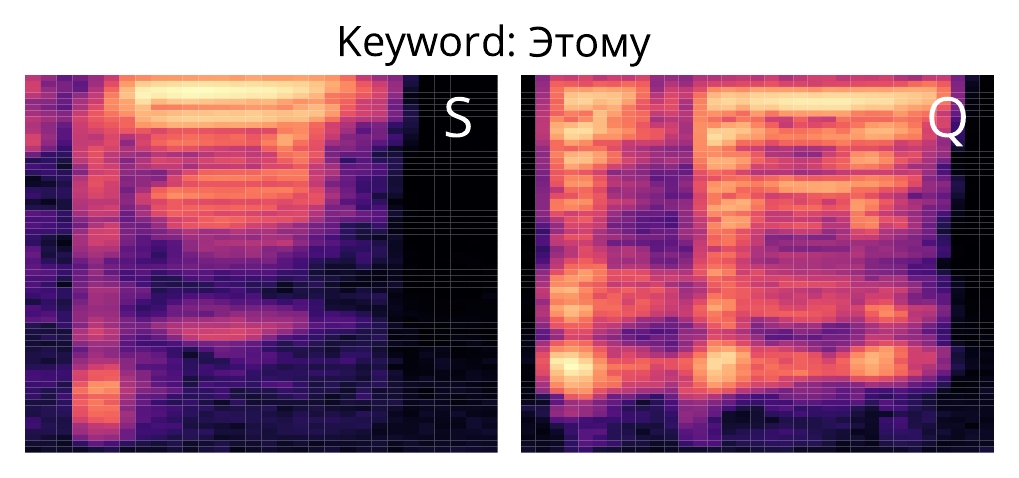}} \\

\subfloat{\includegraphics[width=0.33\textwidth]{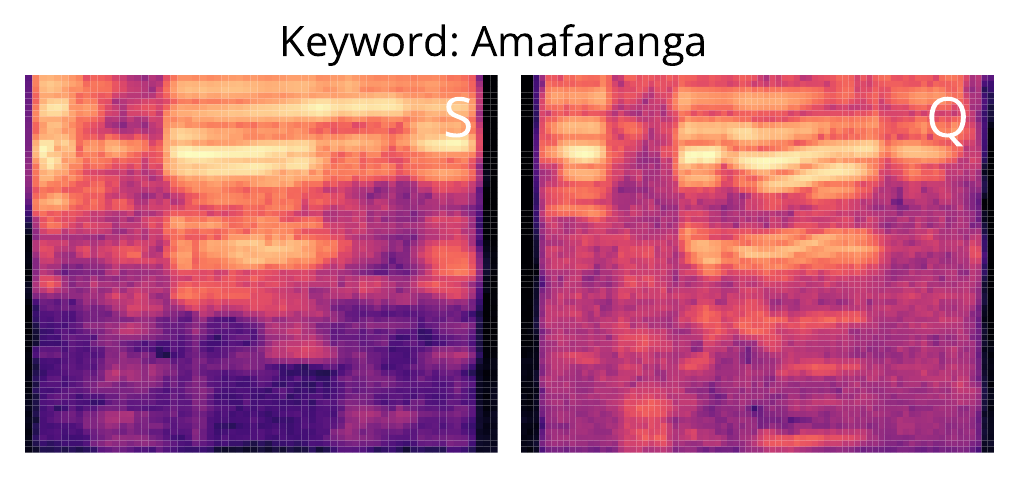}}  
\subfloat{\includegraphics[width=0.33\textwidth]{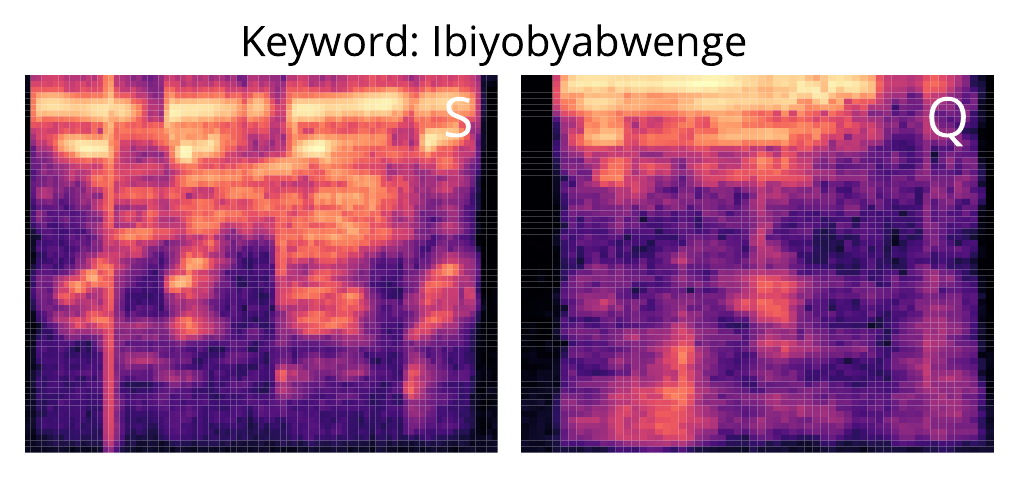}}
\subfloat{\includegraphics[width=0.33\textwidth]{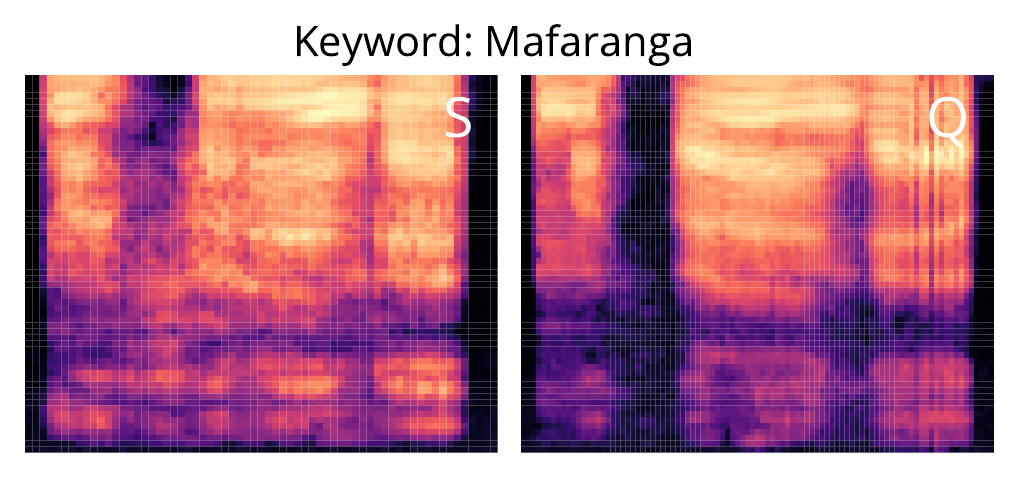}} \\
\caption{We showcase randomly selected query $Q$ and support $S$ set examples from an episode comprising different keywords from five languages (i.e., `en', `de', `es', `ru'. and `rw'). The log-Mel spectrograms are computed for one-second audio clips with $64$ mel-spaced frequency bins.}
\label{fig:lms}
\end{figure}

\subsubsection{Training Procedure}
\label{subsec:tp}
As mentioned before, we use audio resampled at $16$kHz, where each instance comprising a one-second of spoken word. We extract log-Mel Spectrogram from audio clips (see Figure~\ref{fig:lms} for visualization) on the GPU as part of the model for efficient processing of large-scale audio dataset. Specifically, we use window length of $400$, hop length of $160$, and $64$ mel-spaced frequency bins in the range $60$-$7800$Hz. From both considered encoder networks, we extract features from the penultimate layer after applying global average pooling, resulting in a feature of dimension $1280$. We use these features to compute square euclidean distance as a similarity metric between query and support set instances for classification. To generate training episodes, we randomly sample classes (i.e., spoken words) and then sample audio clips corresponding to those clips to form query and support sets. In case of multilingual model training, we combine language-specific datasets to form a larger pool of data and employ earlier explained strategy for episode generation during training phase. This approach makes entire training pipeline highly simplified and result in diverse episodes with data from different languages, which can be seen as implicit multi-task learning~\cite{caruana1998multitask}. We use Adam optimizer with learning rate of $0.001$ to train models for a $100$K episodes except for the `base' multilingual model that is trained for $200$K episodes due to larger combined trained set. In all cases, during training the number of queries are set to $10$, unless stated otherwise. Furthermore, we use ImageNet pretrained weights for initialization as we found them to be resulting in faster convergence than using randomly initialized ones. We believe initialization can be further improved by using a model pretrained on audio dataset, we leave this as a topic for future research.

\subsubsection{Inference Strategies}
\label{subsec:is}
Once few-shot model is learned with episodic data of spoken words as explained in Section~\ref{subsec:tp}, it can be leveraged in a variety of ways to detect audio keywords in-the-wild and to solve other related problems (e.g., spoken language detection). At test-time, user provides query examples (i.e., audio clips) that may contain keywords of interest to be detected along with few support examples. In case of audio clip being longer than one second, one can chunk them into one second clips and treat each resulting clip as one query example. Few-shot model can then compute similarity of query examples with those in the support set and assign labels based on the prototype distance. We note that ProtoNet does not require having similar configuration of query and support set at inference time as leveraged during training. For instance, at training time we can learn a model of $N$-way $K$-shot, where can be $N = 10$ and $K = 5$ but at inference time we can vary both $N$ and $K$. However, note that performance may degrade significantly if $N$ and $K$ both vary considerably than training configuration. Furthermore, in one-shot setting, where at test-time user only provides one support example, we can leverage inference time augmentation~\cite{wang2020few} to generate multiple support examples. 

\subsection{Plug-and-Play Approach and Open Source}
Our approach aims to simplify the deployment of spoken word recognition models in real-world scenarios. In general, developing and maintaining standard supervised models to recognize spoken keywords reliably is challenging, as users' requirements for the spoken words of interest may change over time. Hence, our few-shot model provides a feasible way to keep supporting the detection of novel spoken words using minimal labeled data in the form of support examples. Furthermore, our plug-and-play design enables easy integration into existing systems, requiring minimal modifications, and supports recognition across multiple languages with only a few support set examples. In order to enable other researchers to explore this domain and for developers to easily use it in prospective applications, we also release inference code and model weights, including multilingual and language-specific models. With this release, we are further aiming to facilitate adaptation and extension for novel use cases. Our pre-trained models can be leveraged for transfer learning on low-resource languages. They can enable high levels of accuracy while circumventing the need for extensive training on large datasets, a process that can be both time-consuming and resource-intensive.

\section{Experiments}\label{sec:experiments}
\subsection{Multilingual Spoken Words Dataset}

\begin{figure}[!t]
    \centering
    \includegraphics[width=0.8\textwidth]{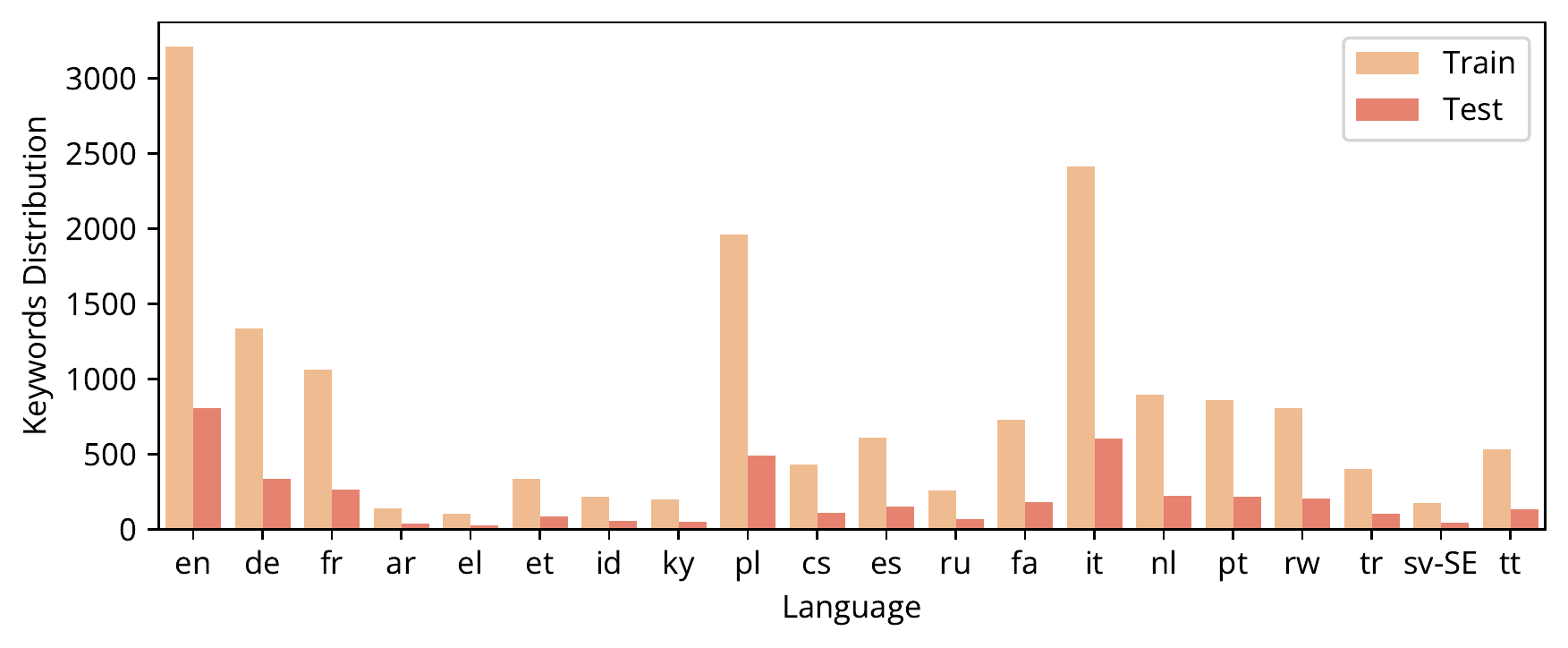}
    \caption{Unique keyword (classes or spoken words) in `train' and `test' sets across considered languages.}
    \label{fig:words_dist}
\end{figure}

\begin{figure}[!t]
    \centering
    \includegraphics[width=0.8\textwidth]{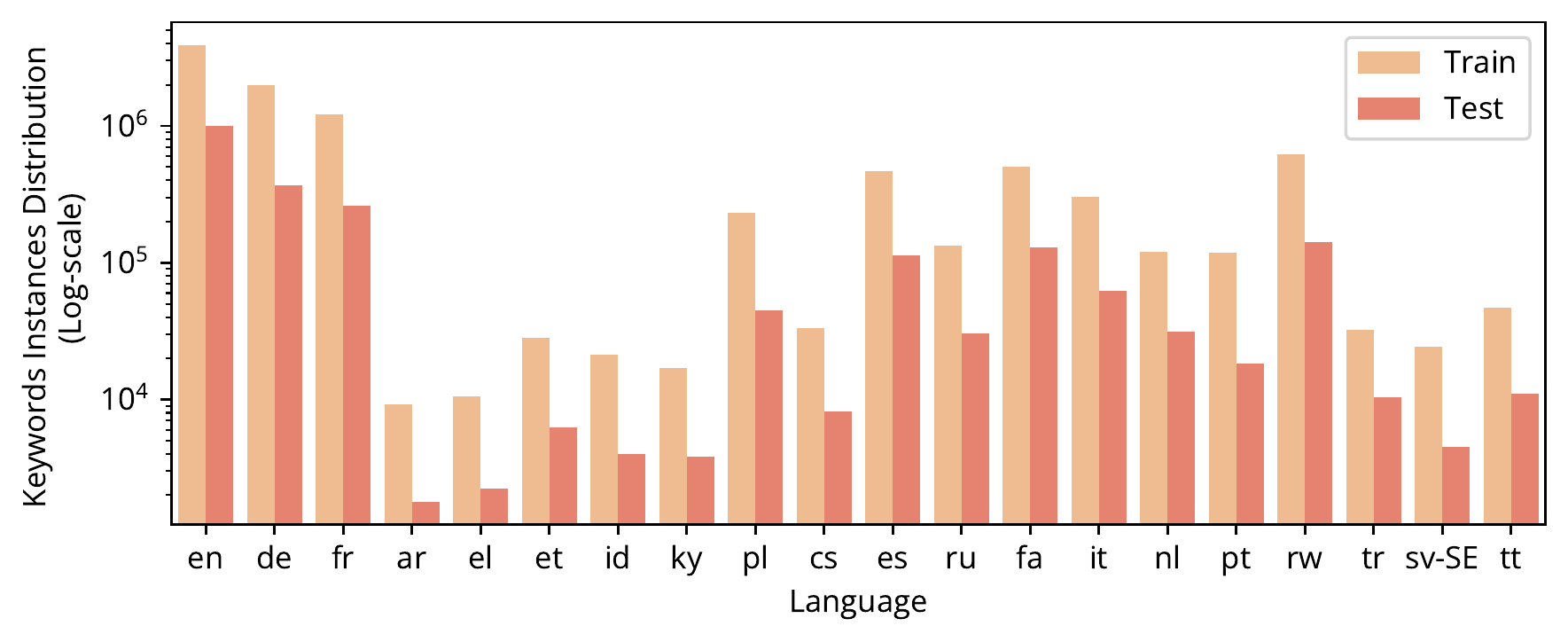}
    \caption{Number of audio clips per language, highlighting resource richness of the spoken language as represented in MSWC.}
    \label{fig:samples_dist}
\end{figure}

We use Multilingual Spoken Words Corpus (MSWC)~\cite{mazumder2021multilingual} for training and evaluation of our few-shot learning models. It is intended for use both in academic research and commercial applications, particularly in the field of keyword spotting. MSWC is a vast and expanding collection of audio recordings containing spoken words in $50$ different languages. The covered languages are spoken by a combined total of more than $5$ billion individuals. Specifically, it contains one-second spoken words for more than $340$K keywords, totaling $23.4$ million examples. It is generated automatically through force alignment on a crowd-sourced speech corpus, Common Voice dataset~\cite{ardila2019common}. Due to large number of keywords and their associated audio clips, it lands as an ideal candidate for episodic training, where, each episode implicitly requires to contain novel classes. For benchmarking purposes the corpus is split by default into train, dev, and test splits of sizes $80\%$,$10\%$, and $10\%$, respectively. The clips in the dataset are are opus-compressed using a single channel with a $48$KHz sample rate. In our work, as a preprocessing we convert all audio files to a wav format resampled at $16$KHz. We focus on $20$ high and medium resource languages from the MSWC corpus. We filter keywords that has less than $200$ audio clips for high-resource languages: `en', `de',  `es', `fr', `fa', `ru', and `rw'. Likewise, we also remove keywords for rest of the considered languages if number of audio clips are less than $25$. Note that few-shot models are evaluated with a novel (or disjoint) set of classes that the model is not exposed to during training phase. To this end, we initially combine keywords `train' and `dev' sets provided in the corpus and then perform splits with ratio of $80:20$ based on keywords such that there is no overlap in keywords of train and test splits. We provide unique keyword distribution per language in Figure~\ref{fig:words_dist}, where, English language has largest number of keywords. Furthermore, we show audio clips (one-second each) in our generated splits of the data in Figure~\ref{fig:samples_dist}. Our generated dataset for few-shot learning and evaluation collectively contains more than $12$ million one-second audio clips with $16,665$ and $4,174$ unique keywords in training and testing sets, respectively.  

\subsection{Evaluation Protocol}
We perform evaluation of few-shot models on unseen spoken words (or classes) in our test set. We use $1000$ episodes for testing, where each episode comprises $15$ queries unless stated otherwise. Note that, we do systematically vary number of ways and support examples that we mention when discussing respective results in subsequent section. The testing episodes are constructed in a manner that mirrors the training phase. Specifically, we randomly sample class labels and, for each class, we sample a specified number of support and query instances. During testing, we identify spoken keywords in a test recording by assigning a label to a query example based on the closest prototype distance. We report the average accuracy over all the episodes under all experimental conditions.

\subsection{Results}

\begin{figure}[!t]
\subfloat{\includegraphics[width=0.9\textwidth]{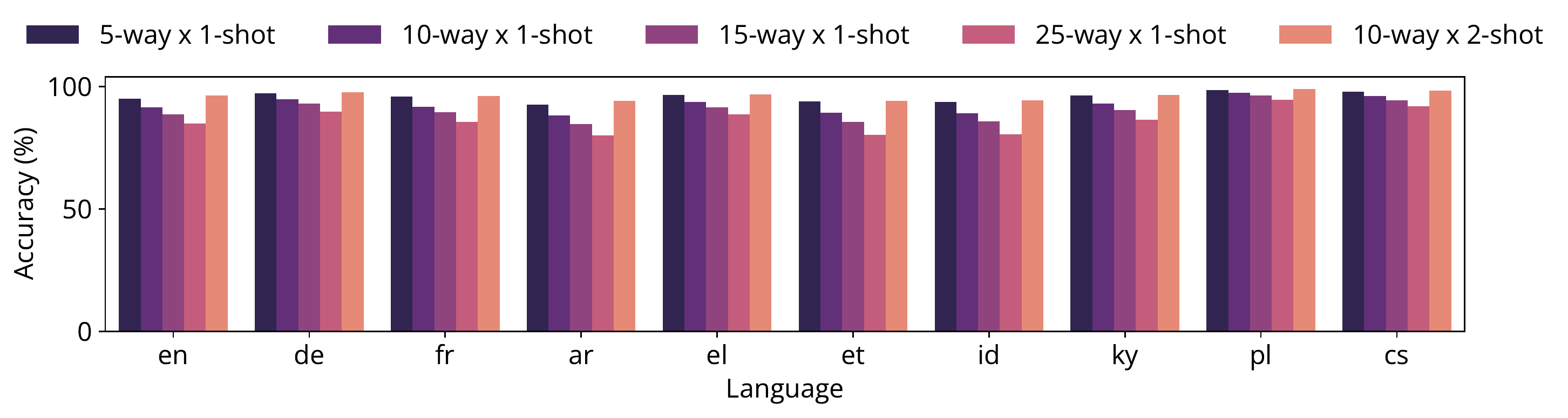}}  \\
\subfloat{\includegraphics[width=0.9\textwidth]{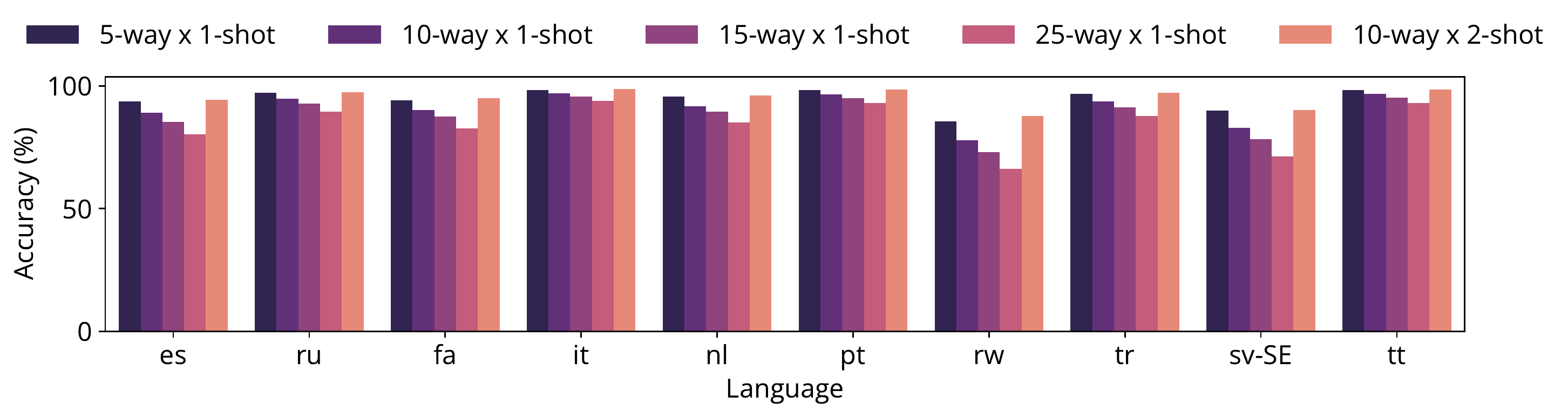}}
\caption{Performance evaluation of multilingual `base' model trained on data from considered $20$ languages. The few-shot model is trained for a $10$-way ($N$) and $5$-shot ($K$) classification task. At inference-time we vary both $N$ and $K$ to demonstrate generalization capability even when only support example is provided.}
\label{fig:base_ml}
\end{figure}

\subsubsection*{Power of a Single Multilingual Model:} We begin with examining the generalization capabilities of our multilingual `base' model (i.e., EfficientNet-v2) across a range of few-shot evaluation scenarios. Specifically, we systematically vary the number of ways ($N$) and the number of support examples per class ($K$) to determine how the model's performance generalizes to unseen words on considered languages. In our experiment, we first train the base model for $10$-way $\times$ $5$-shot classification using only words from the training sets of each language, we then evaluate the learned model on held-out words. We present these results in Figure~\ref{fig:base_ml}. For $5$-way, $10$-way, and $15$-way classification with a single support example per class, we find that the model achieves superior performance above $80\%$ accuracy, reaching above $90\%$ accuracy for $5$-way classification in the majority of cases. Even for more challenging $25$-way classification, the model produces accurate recognition on resource-rich languages like English and German. Furthermore, increasing the number of support examples to $2$ for $10$-way classification results in consistently superior detection rates. However, for languages like Kinyarwanda (rw) and Swedish (sv-SE), $25$-way $\times$ $1$-shot classification achieves relatively lower accuracies of $66\%$ and $71\%$, respectively. Based on this, we increase the number of support examples to five and this instead yields substantial improvements to $87.6\%$ accuracy for `rw' and $88.9\%$ accuracy for `sv-SE', demonstrating the benefits of increasing support set size even by just a few examples for languages manifesting varied acoustic and structural characteristics. Overall, these results highlight the strong generalization capabilities of our multilingual base model, especially given limited data. The model's performance suggests promising capabilities for few-shot recognition of spoken words with a single model, particularly with increasing support examples to help mitigate the challenges of low-resource languages.

\begin{figure}[!htbp]
\includegraphics[width=0.75\textwidth]{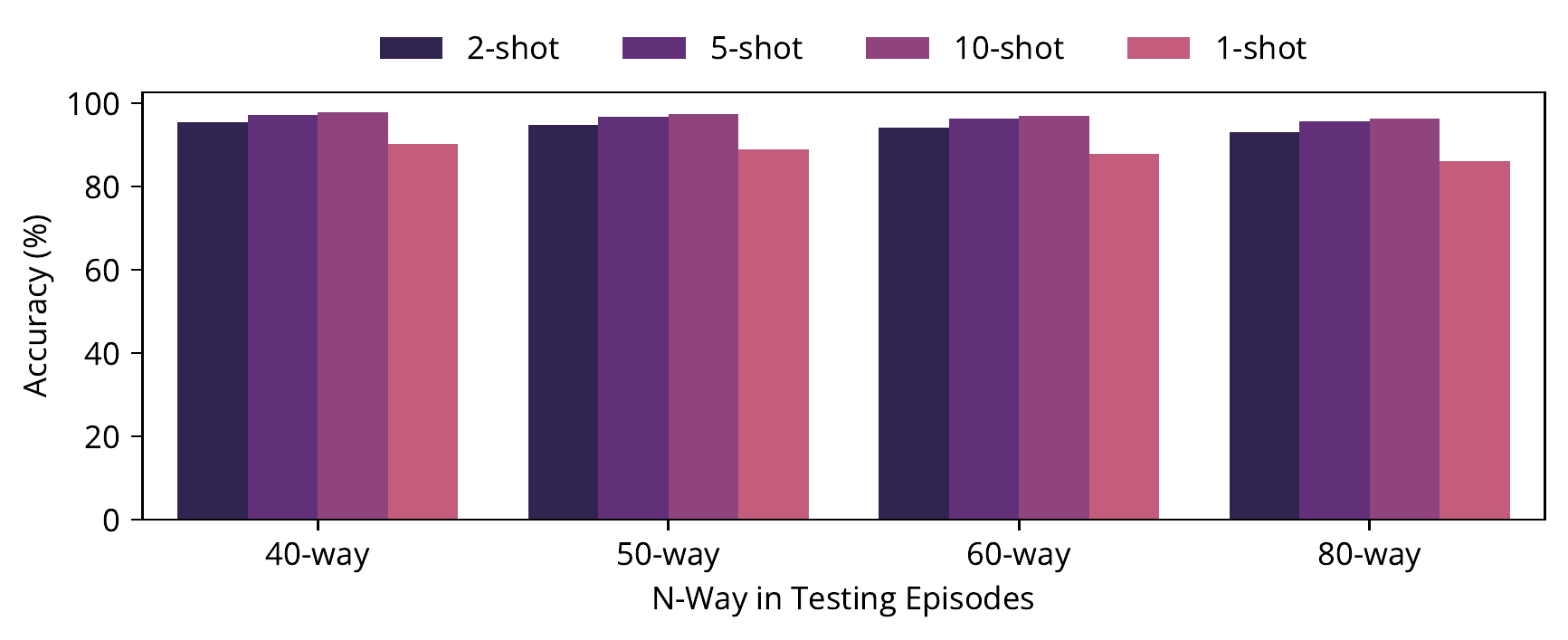}
\caption{Result of few-shot English language-specific `base' model learned with $30$-way and $5$-shot task. We aim to highlight that at test-time the same model can handle large-number of classification tasks with varying support set examples.}
\label{fig:base_en}
\end{figure}   

\subsubsection* {Accurate Classification with Increasing Classes:} Subsequently, we conduct an empirical evaluation of our `base' model trained only on English language spoken words. We train a few-shot model on a $30$-way $\times$ $5$-shot task and evaluate them under varying settings of $N$ and $K$ during testing. Specifically, we scale the $N$-way (number of classes) classification task from $40$, to $50$, to $60$, and ultimately to $80$ classes, which indicates that once a model is trained, it can differentiate a spoken word from a group of other words even when the number of words is increased but the number of support examples is few. In Figure~\ref{fig:base_en} we provide these results, beginning with when $N$ is set to $40$ and one have a single support example per class, in that case the model achieves $90\%$ accuracy, which declines by merely four percentage points to $86\%$ for $80$-way classification. Upon increasing the number of support examples from one to ten, we observe a gradual increase in detection rates across all considered experimental conditions. With five support examples, the accuracy on the more challenging $60$-and $80$-class tasks reaches $96\%$ and $95\%$, respectively.

\begin{figure}[!htbp]
\subfloat{\includegraphics[width=0.9\textwidth]{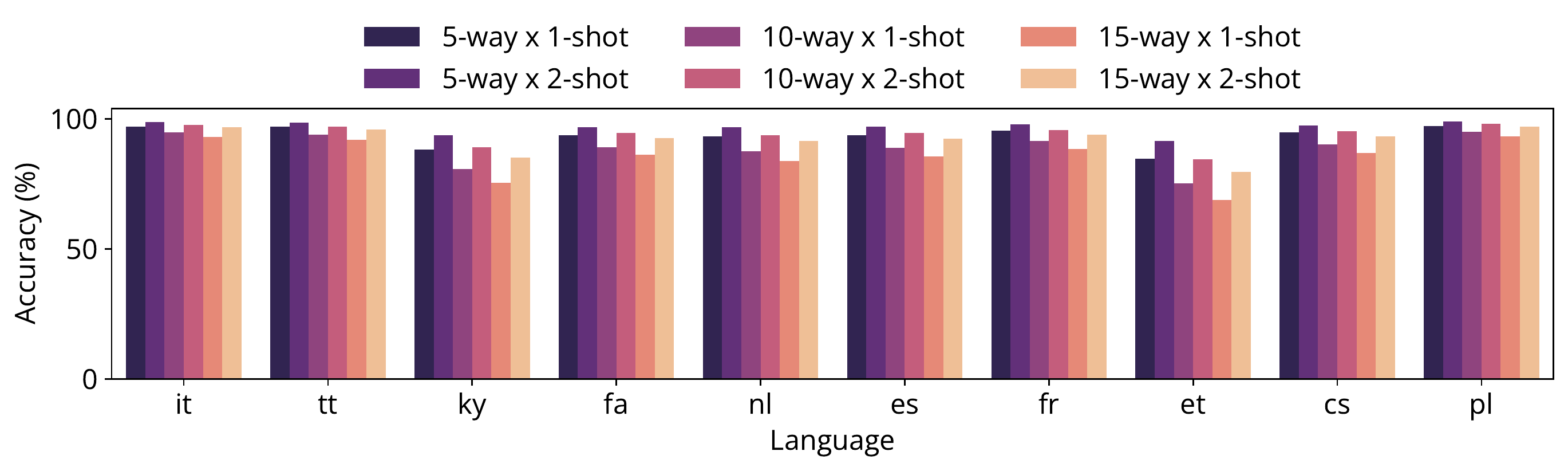}}  \\
\subfloat{\includegraphics[width=0.9\textwidth]{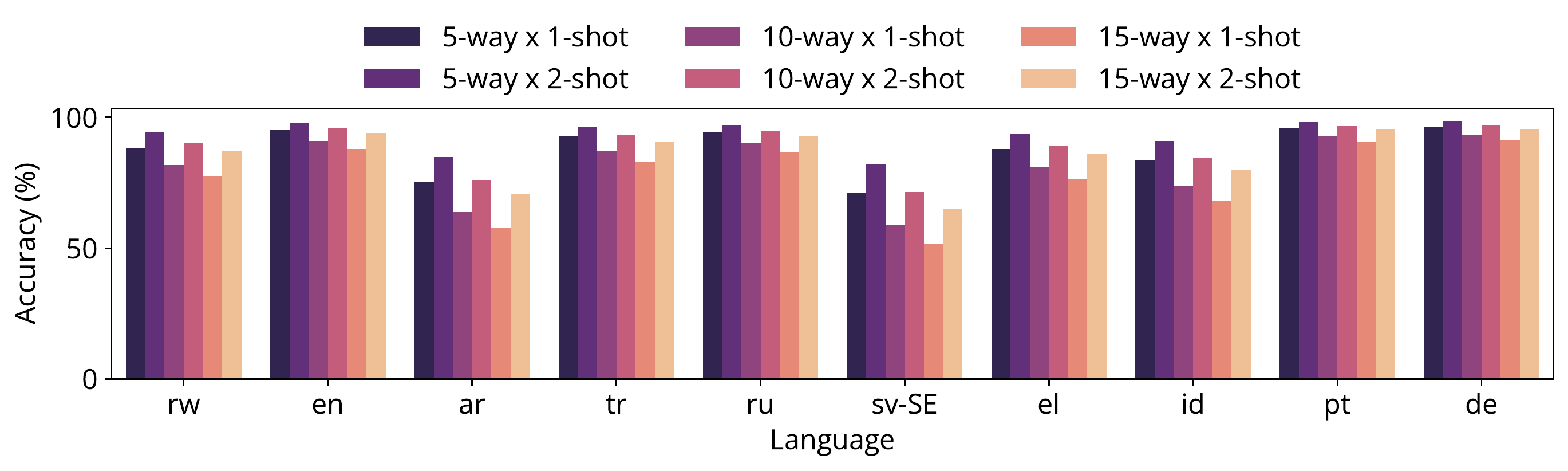}}
\caption{Language-specific `small' models' evaluation under varying number of query and support set examples.}
\label{fig:small_la}
\end{figure}

\subsubsection*{Small but Mighty Language-Specific Few-Shot Models:} In Figure~\ref{fig:small_la}, we present results of our `small' language-specific models (i.e., TinyNet-E) trained and evaluated independently for each of the considered languages. We develop these few-shot models with smaller memory footprint and demand for computational resources as a generalizable model can be relevant for inference on edge and wearable devices. These models are trained in a $10$-way $\times$ $5$-shot recognition task. We evaluate them on varying number of ways and shots. We observe that the `smaller' models for resource-rich languages generalize exceptionally well; for instance, for Italian (`it'), English (`en'), German (`de'), and Spanish (`es'), the performance for $5$-way task with one support example the accuracy is above $95\%$. Even when scaling the classification task to differentiate between $15$ spoken words with $2$ support examples, performance remains above $80\%$ in most of the cases. However, for other languages where dataset size is relatively small, performance is low. This phenomenon is apparent for Arabic (`ar'), Swedish (`sv-SE'), Indonesian (`id'), and Estonian (`et') when support examples are lower and number of ways are greater. Note that we evaluate our models in a highly data-scarce setting, i.e., one or two support set examples are available. However, when we scale support examples to $5$ for $10$-way task in Indonesian (`id') language accuracy reaches to $90\%$.  We see the same effect in Arabic (`ar'), where accuracy increase from $63\%$ to $85\%$ by only scaling support examples from one to five. We conclude based on these results that smaller model can also perform well on unseen spoken words given that either the training episodic data is large enough or, at inference time, more support examples are provided. 

\begin{figure}[!t]
    \centering
    \includegraphics[width=0.75\textwidth]{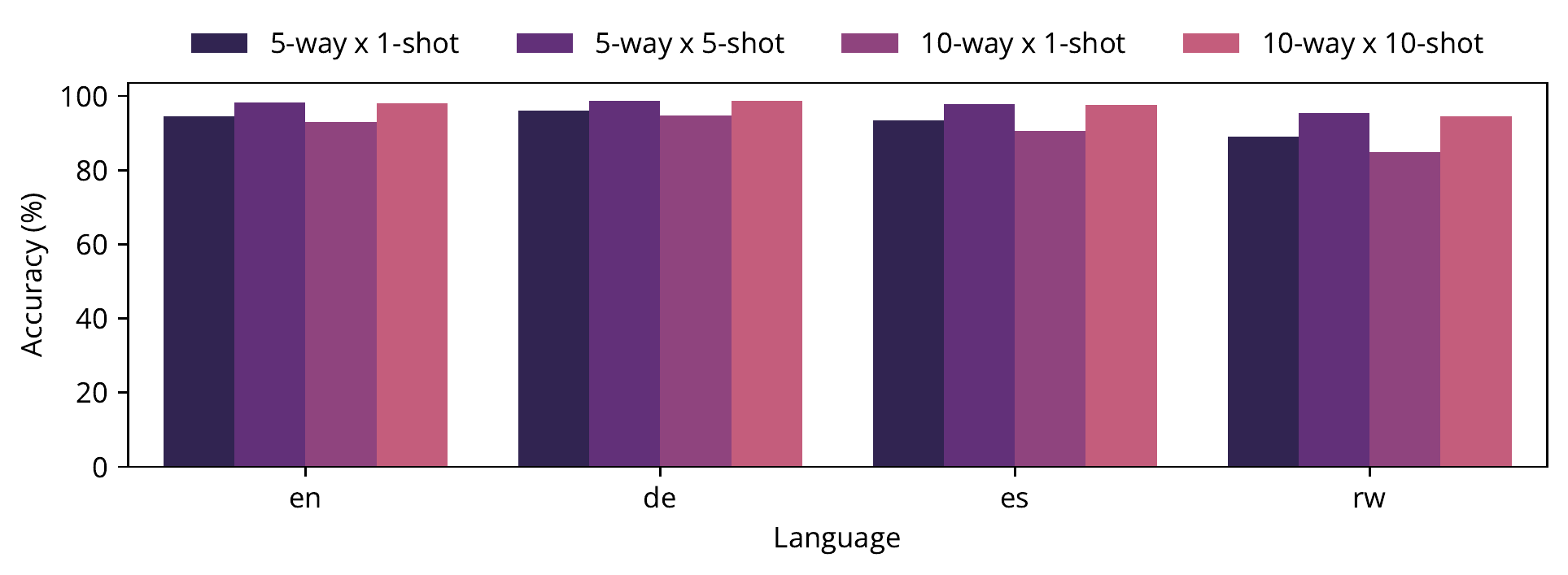}
    \caption{Experimental results of the `small' model as few-shot learner on four resource-rich languages for varying $N$ and $K$ in training and testing setting, i.e., model trained for a $5$-way $\times$ $1$-shot task is evaluated with a same configuration at inference-time.} 
    \label{fig:small_fl}
\end{figure}

\begin{figure}[!htbp]
    \centering
    \includegraphics[width=0.65\textwidth]{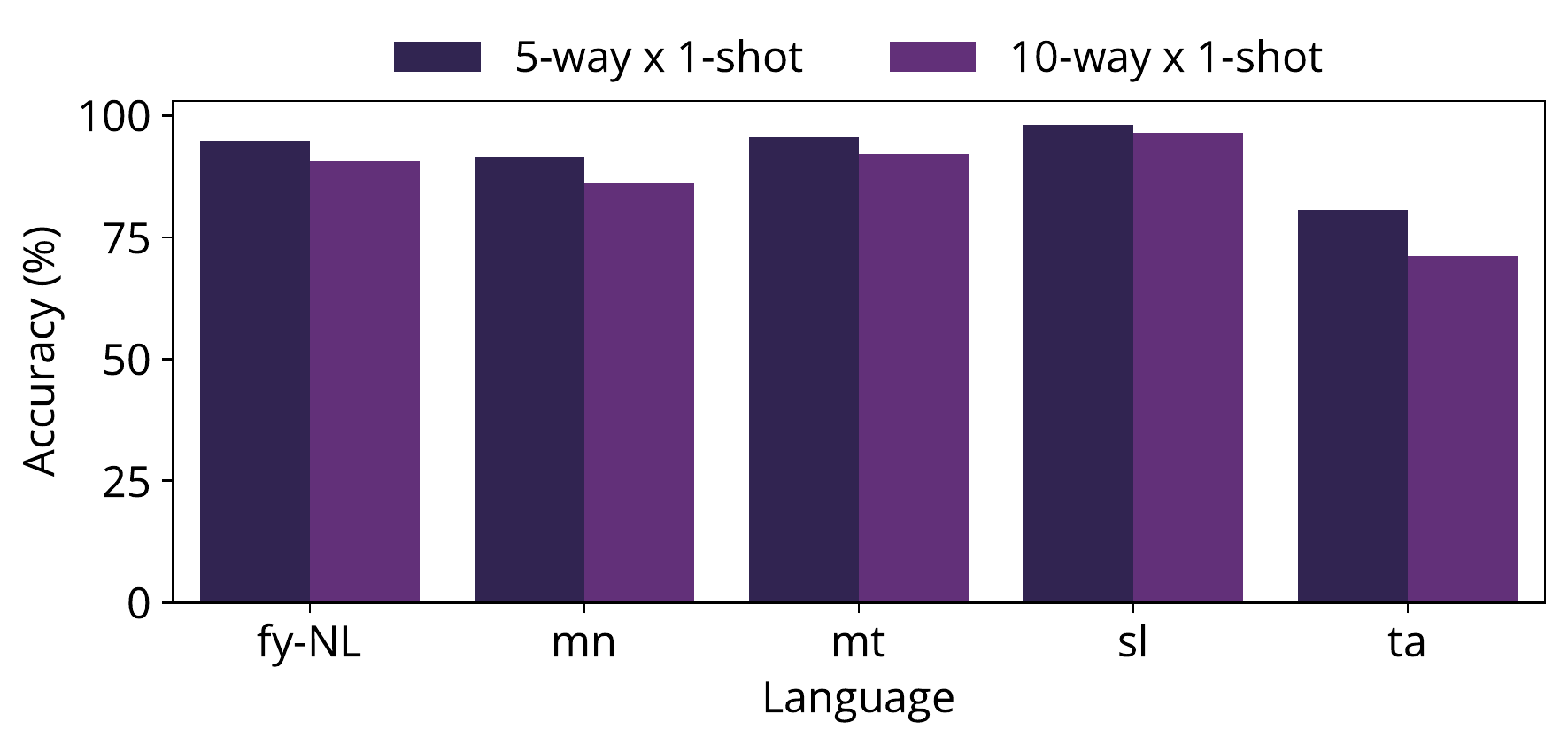}
    \caption{Results on $5$ unseen low-resource languages in MSWC, namely Frisian, Mongolian, Maltese, Slovenian, and Tamil. We evaluate `base' multilingual model in a one-shot manner to show generalization power of our few-shot model.} 
    \label{fig:base_unseen_l}
\end{figure}

\subsubsection*{Varying Few-Shot Parameters:} To further evaluate the generalization capability of our few-shot models trained on resource-rich languages within the context of the MSWC, we train and test our `small' models under identical configurations, specifically retaining the same $N$-way and $K$-shot parameters. As evidenced in Figure \ref{fig:small_fl}, we observe notable improvements in model performance, particularly for 'rw', where accuracy reaches $84\%$ for the $10$-way task with a single support example, representing an increase of three percentage points relative to the evaluation setup and results presented in Figure \ref{fig:small_la}. While performance improvements for other languages are more modest, this suggests an intriguing phenomenon whereby few-shot models can recognize spoken words with high accuracy even when $N$ and $K$ are varied during the test phase, a valuable characteristic from a real-world deployment perspective.

\begin{figure}[!htbp]
    \centering
    \includegraphics[width=0.65\textwidth]{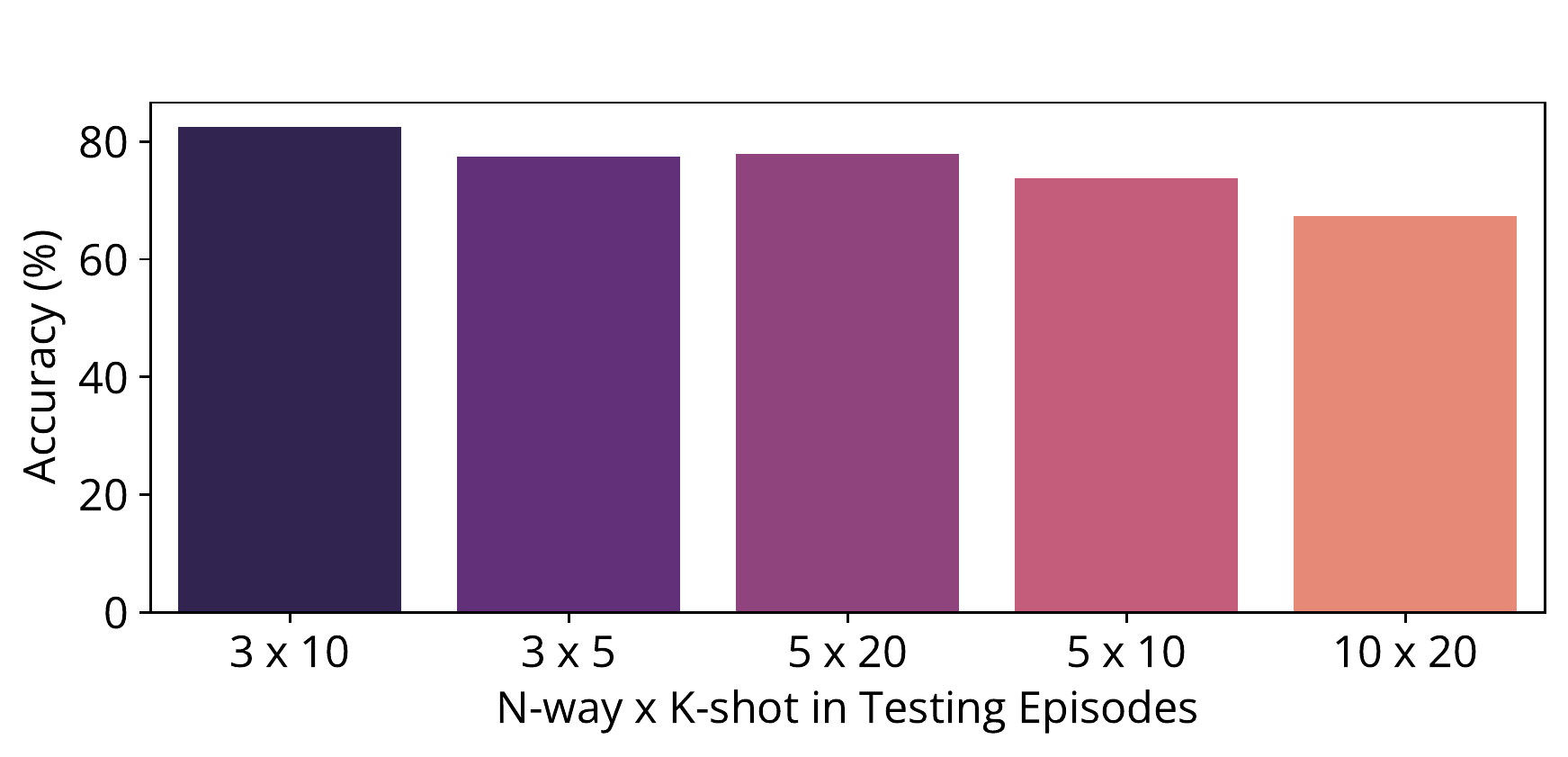}
    \caption{Transfer evaluation results on FLEURS~\cite{fleurs2022arxiv} language identification task.} 
    \label{fig:fleurs}
\end{figure}

\begin{figure}[!t]
\subfloat[English]{\includegraphics[width=0.5\textwidth]{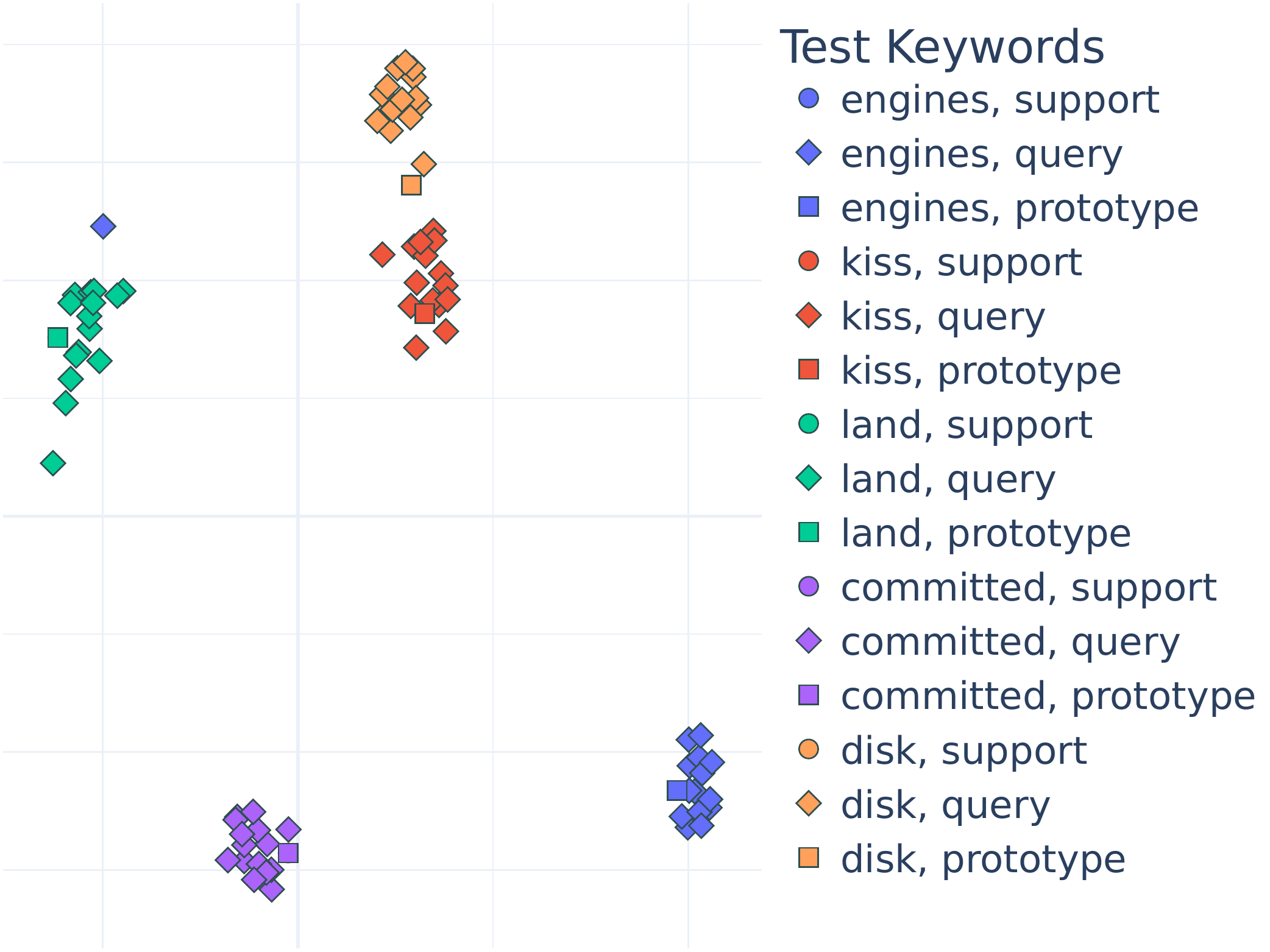}} 
\subfloat[Arabic]{\includegraphics[width=0.5\textwidth]{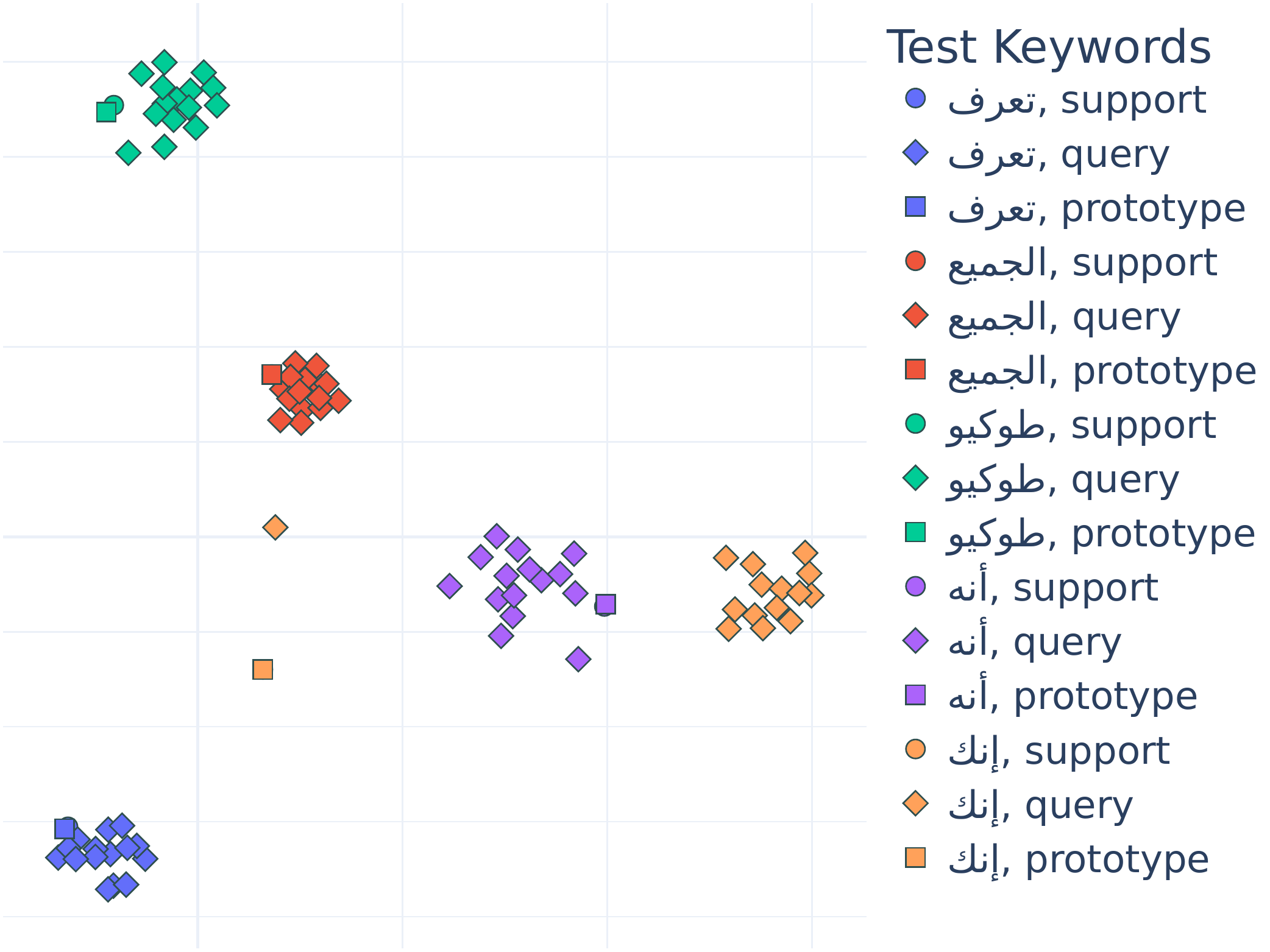}} \\
\subfloat[Kyrgyz]{\includegraphics[width=0.5\textwidth]{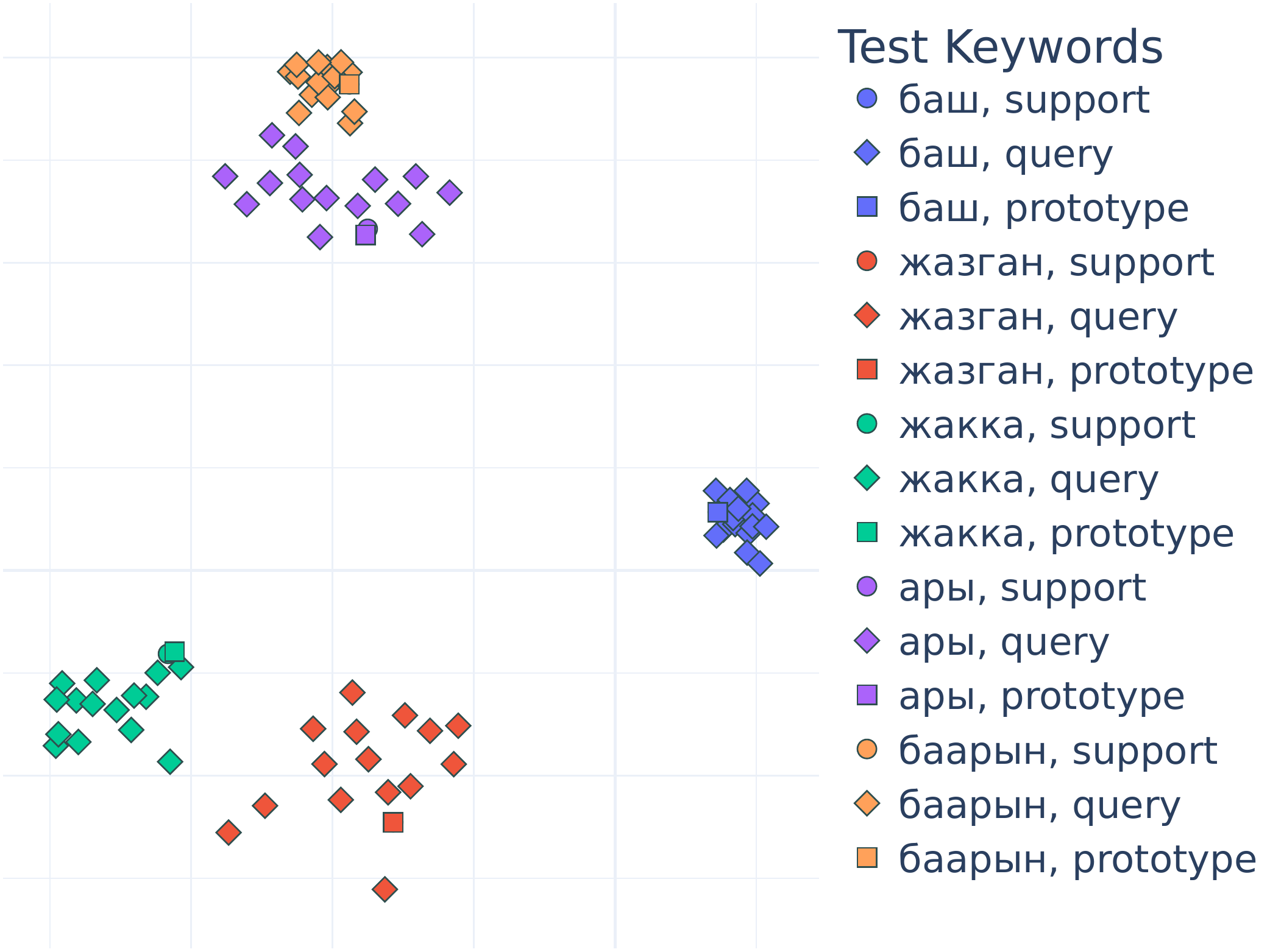}}
\subfloat[Kinyarwanda]{\includegraphics[width=0.5\textwidth]{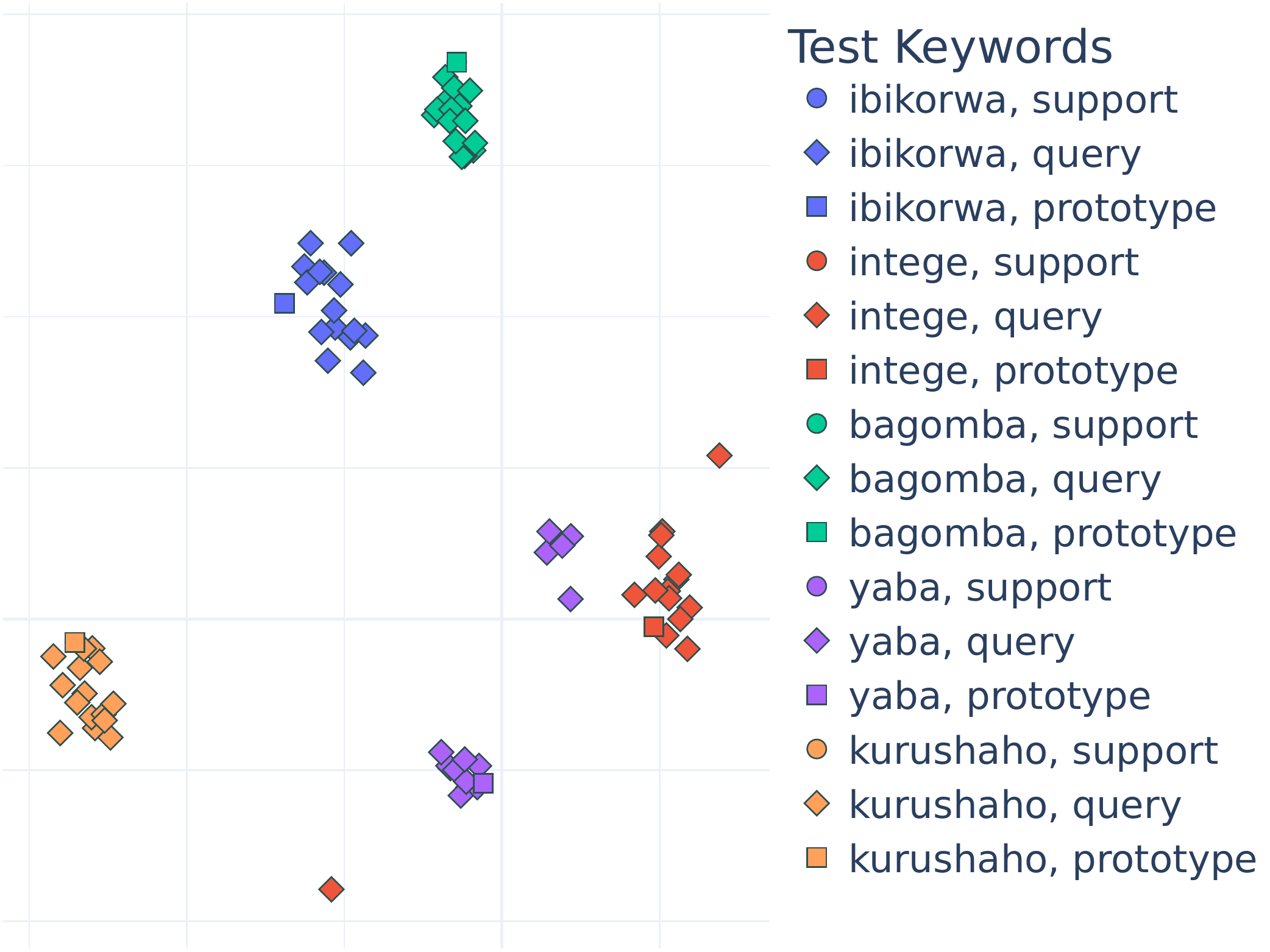}} 
\caption{t-SNE visualization of the query and support set examples along with computed prototypes per class. We use a multilingual `base' model to extract embeddings for $5$-way $\times$ $1$-shot task. For the sake of simplicity, we only visualize for a single episode across four languages. Better visualized in color and when zoomed in.}
\label{fig:tsne}
\end{figure}

\subsubsection*{Cross-lingual Generalization:} We also investigate the generalization capability of our `base' multilingual model in recognizing spoken words in previously unseen languages, that is, languages on which the model has not been trained. To this end, we select five low-resource languages: Frisian (`fy-NL'), Mongolian (`mn'), Maltese (`mt'), Slovenian (`sl'), and Tamil (`ta'). We filter out keywords with fewer than $15$ audio clips and evaluate the model in a challenging one-shot setting. The model performs robustly out of the box without retraining on these novel languages. Specifically, it achieves above $91\%$ accuracy except for Tamil, where the model achieves $80\%$ accuracy in a $5$-way $\times$ $1$-shot task. These results suggest the model's ability to capture linguistic features that extend beyond specific languages. Though, a slight degradation in accuracy on Tamil suggests that the model may struggle with languages more distant from the languages it was trained on. Further work is needed to improve generalization to such linguistically diverse languages.

\subsubsection*{Transfer Evaluation on Language Identification Task:} 

For completeness, we also evaluate our `base' multilingual model trained in a few-shot manner in a transfer setting on a FLEURS~\cite{fleurs2022arxiv} language identification task, comprising $102$ languages in total. The task involves recognizing spoken language given an audio clip. We use test set instances from the dataset with each audio clip trimmed or padded to be of $10$ seconds in length. We do not change the model architecture nor perform any training steps to show the generalization power of our model. We evaluate the model in a $3,5,$ and $10$ way setting with a number of support examples in the range of $5, 10$, and $20$. We notice our model achieves above $82\%$ accuracy in the case of a $3$-way $\times$ $10$-shot task and above $73\%$ in the rest of the settings except for $10$-way task. The drop in accuracy, in this case, is expected given the increased complexity of distinguishing between $10$ languages with limited examples. Overall, these results highlight the potential of our multilingual few-shot approach to achieve solid performance with limited data in unseen languages and spoken words.

\subsubsection*{Visual Representation Analysis with t-SNE:} Moreover, in Figure~\ref{fig:tsne}, we provide the t-SNE~\cite{van2008visualizing} embedding from a `base' model on four languages (i.e., `en', `ar', `kr', and `rw') to demonstrate the disentanglement of the learned representations based on different spoken words. We use $5$-way $\times$ $1$-shot episodes for generating embeddings and apply t-SNE to generate $2$D projections. Despite being exposed to merely a singular support exemplar per class, our model generates clearly demarcated clusters corresponding to various spoken words across languages. These results further indicate that our model is capable of efficiently learning representations that disambiguate acoustic characteristic of spoken words across languages with limited labeled data at test-time, thus demonstrating promise for cross-lingual few-shot recognition.

\subsubsection*{Resource Consumption Profiling:} 
\begin{figure}[!t]
    \centering
    \includegraphics[width=0.45\textwidth]{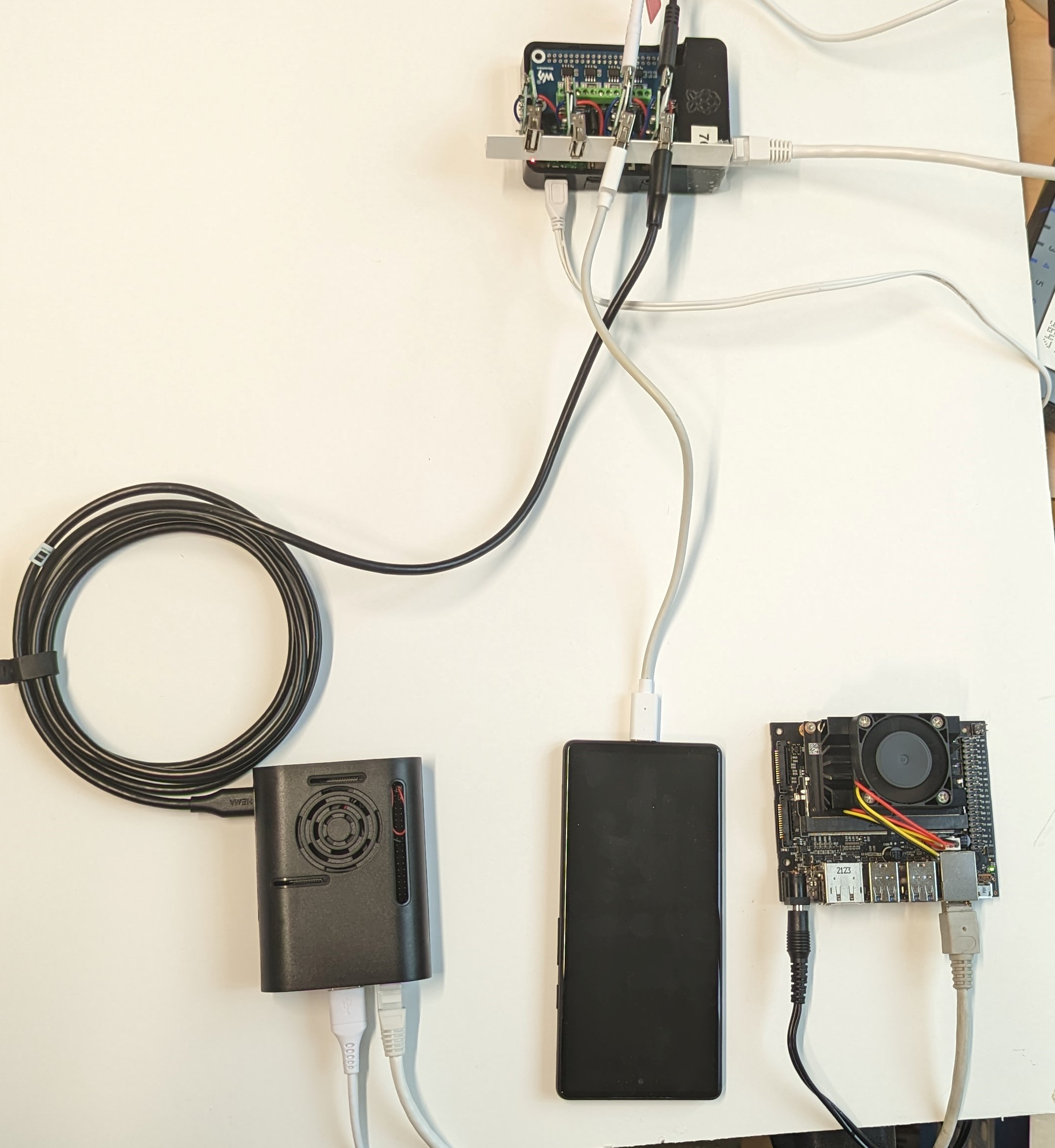}
    \caption{Embedded platforms setup for energy and latency consumption estimation of our models on variety of devices.} 
    \label{fig:gadgets}
\end{figure}

For each of the encoders (`small' and `base') discussed in the Section~\ref{sec:methodology}, we examine their runtime resource demands, e.g., computation and energy across three embedded platforms, namely NVIDIA Jetson Nano (4GB), Raspberry Pi 4, and Pixel 7 (see Figure~\ref{fig:gadgets} for the setup). We conduct this analysis using only encoder to keep the setup simple and also because encoder models are the central components comprising of large number of learnable of parameters. We chose these platforms based on their popularity in the market and their suitability for running machine learning workloads on edge devices.

To test our system, we utilized PyTorch benchmarking modules for edge\footnote{\url{https://pytorch.org/tutorials/recipes/recipes/benchmark.html}} and mobile\footnote{\url{https://pytorch.org/tutorials/recipes/mobile_perf.html}} devices, while both versions of our networks are evaluated: a full precision (float32) and a quantized models optimized for mobile devices. As the PyTorch benchmarking modules for mobile devices only allows for quantized models to be evaluated on devices, we only performed experiments with the  quantized variant on Pixel 7. To determine the energy consumption across platforms, we utilized an external equipment (a Raspberry Pi 4 with a Current/Power Monitor HAT) that monitors the energy usage of each device individually using a sampling resistor of $0.1$Ohm. In case of Pixel 7, we have deactivated the battery to eliminate the possibility of interfering with our measurements. Table~\ref{tab:energy_latency} illustrates the average running time (in milli-seconds) and energy consumption (in mJ) over $200$ inference runs of the deep models observed on all three platforms. 

\begin{table}[t]
    \begin{minipage}[b]{0.45\linewidth}
    \centering
    \begin{tabular}{@{}llcc@{}}
        \toprule
        \textbf{}  & \textbf{Device}    & \textbf{Small} & \textbf{Base} \\ \midrule
        \multirow{2}{*}{Standard}  & Raspberry Pi & 240.17         & 4480          \\
        & Jetson Nano  & 35.12          & 1954          \\ \midrule
        \multirow{3}{*}{Quantized} & Pixel        & 5.12           & 78.07         \\
        & Raspberry Pi &    6.92            &      107.13         \\
        & Jetson Nano  &    7.32            &      101.37         \\ \bottomrule
    \end{tabular}%
    
    \caption*{Latency (milli-seconds)}
  \end{minipage}
  \hspace{0.05\linewidth}
  \begin{minipage}[b]{0.45\linewidth}
    \centering
    \begin{tabular}{@{}llcc@{}}
        \toprule
        \textbf{} & \textbf{Device}    & \textbf{Small} & \textbf{Base} \\ \midrule
        \multirow{2}{*}{Standard}  & Raspberry Pi & 137.82         & 2011.44       \\
        & Jetson Nano  & 133.53         & 1227.24       \\ \midrule
        \multirow{3}{*}{Quantized} & Pixel        & 3.55           & 39.71         \\
        & Raspberry Pi &   4.15             &     71.75          \\
        & Jetson Nano  &   4.92             &     64.37          \\ \bottomrule
    \end{tabular}%

    \caption*{Energy (milli-Joule)}
  \end{minipage}
  \caption{Average energy and latency benchmarking of the considered encoders, i.e., TinyNet-E and EfficientNet-v2 (M) across different devices. Note that for the sake of simplicity, we only use encoder networks discarding other components (like e.g., log-Mel Spectrogram layer) for computing these metrics. Pixel 7 only supports quantized models.}
  \label{tab:energy_latency}
\end{table}

We note that our models are efficient across various hardware devices like the Raspberry Pi, Jetson Nano and Google Pixel. On the Raspberry Pi, the standard EfficientNet-v2 (M) model has a latency of $4480$ ms and energy usage of $2011.44$ mJ. The `small' model is more efficient with 240.17 ms latency and 137.82 mJ energy usage but with a trade-off in terms of generalization. Further, applying quantization techniques to these models leads to significant improvements in efficiency. The quantized `base' encoder on the Raspberry Pi has a latency of just $107.13$ ms and $71.75$ mJ energy usage - demonstrating $19$x and $44$x improvements over the standard model. Similarly, the quantized `small' encoder on the Jetson Nano has $7.32$ ms latency and $4.92$ mJ energy usage, showing $4$x and $33$x gains over the standard model. In summary, while our standard encoders can operate efficiently on resource-constrained devices, quantization enables greater speeds and energy efficiency. These optimized models enable low-power AI applications on edge hardware. While our results demonstrate the potential to deploy few-shot models in a computationally viable manner even on extremely resource-limited computing substrates.

\subsection{Potential Applications and Use-cases}
We believe our contributions in this paper can be impactful in many areas. To that end, we conclude with several example uses that demonstrate the potential of our system which span a range of contexts, platforms and application domains.

\noindent \textbf{Virtual personal Assistants:} With the ability to recognize new user-specific commands with just a few examples, our system could make virtual personal assistants like Alexa or Siri even more helpful and responsive. Users could easily teach their assistants to understand new words or phrases and customize them to their specific needs.

\noindent \textbf{Smart Home Automation:} Few-shot spoken word recognizer could be used to control smart IoT devices with voice commands. By recognizing new commands with just a handful of examples, users could easily personalize their smart home systems to recognize and respond to novel voice commands, making home automation even more seamless.

\noindent \textbf{Language Learning:} Few-shot learner could be used to create personalized language learning tools that adapt to the individual needs of each learner. These systems could recognize new words and phrases with just a minimal examples and provide immediate feedback and pronunciation tips to learners, making language learning more efficient and effective.

\noindent \textbf{Speech Therapy:} It also has the potential to serve as an effective tool in speech therapy aimed at enhancing the speech ability of individuals suffering from speech impairments. Through their ability to detect incorrect pronunciation, our system can provide prompt feedback and support to assist patients in refining their pronunciation and improving their overall fluency of speech.

\noindent \textbf{Accessibility:} It can also be used as a means to improve accessibility in assistive applications for people with disabilities as customization would only require a few examples from the individual, making it easier to personalize the system for specific accessibility requirements.

\noindent \textbf{Informatics:}  In the realm of informatics, our system can improve the efficiency of logging and searching for specific words or phrases in lengthy audio clips. The technology's practical applications extend to numerous fields, including journalism, law enforcement, or customer service, where it is essential to quickly find and review specific information in audio recordings to automate the operations.  %

\noindent \textbf{Gaming:} To create more immersive and interactive gaming experiences. Players could use voice commands to control game characters or make decisions. The system's capability to recognize novel voice commands with only a few support examples would enable game-play to become more personalized, thus augmenting the level of engagement for the user.

\section{Related Work} \label{sec:related_work}
There has been extensive research in the area of few-shot learning, which aims to learn new concepts from just a few examples. As this paper explores few-shot learning approach for spoken word detection, it is important to review relevant prior work in this domain. Key approaches to few-shot learning developed techniques that attempted to extract features or learn update rules that could generalize to new concepts from limited data. Several approaches have been proposed to this end over the years, mainly along meta-learning, generative modeling, and metric learning. Meta-learning based methods learn an algorithm or model that can quickly adapt to new tasks with few examples. This includes learning to learn a classifier~\cite{finn2017model}, learning to propagate labels to unlabelled examples~\cite{ravi2017optimization}, and learning to compare examples for similarity via memory-augmented networks~\cite{santoro2016meta}. Generative approaches aim to generate synthetic yet plausible examples for new categories to augment the training data. These include learning to generate images~\cite{schwartz2018encoder} and data augmentation~\cite{chen2019image, schwartz2018delta}. Metric learning based methods~\cite{snell2017prototypical, sung2018learning} learn a distance metric to classify similar and dissimilar examples, even with few examples. In this realm, seminal work has explored the use of prototypes-based nearest neighbor search~\cite{snell2017prototypical} and attention mechanisms~\cite{vinyals2016matching} to enable models to leverage prior knowledge when learning novel concepts. These methods have achieved promising results on benchmark few-shot classification tasks. Due to the great success and simplicity of distance-based methods on not just vision but importantly for sound recognition tasks~\cite{chou2019learning,wang2020few,shi2020few}, we leverage Prototypical Network~\cite{snell2017prototypical} for developing few-shot keyword spotting system. 

The early work on spoken word detection has explored various deep learning approaches, including different neural architectures as convolutional and recurrent networks~\cite{chen2014small,sainath2015convolutional,chen2015query}. While these methods have achieved compact model sizes, their training requires thousands of samples of the target keyword. Other approaches for keyword spotting in low-resource languages employ techniques such as multilingual bottleneck features~\cite{menon2018feature}, utilizing speech synthesis~\cite{lin2020training}. Similarly, ~\cite{san2021leveraging} design a spoken term search system on 10 languages to go beyond monolingual spotting. More recently,~\cite{mazumder2021few} explores transfer learning to fine-tune a model with minimal labeled data to improve label efficiency. In comparison, our work propose a few-shot learning system (PLiX) that can recognize spoken keywords in a broad set of languages without retraining to recognize unseen words (classes). We build a multilingual acoustic model that learns shared representations across languages during meta training phase as opposed to transfer~\cite{mazumder2021few}, enabling the model to generalize to possibly new languages and keywords. Further, our method is general-purpose as it does not require language-specific information or resources (while only using raw audio for learning), achieving strong performance with just a few support samples per keyword. To the best of our knowledge, the proposed PLiX system is the first plug-and-play few-shot spoken word detection system that is trained on millions of audio clips and scales to thousands of keywords. Compared to prior approaches, PLiX achieves superior scalability in terms of the number of keywords and languages supported, enabling its applicability to a wide range of real-world scenarios.

\section{Conclusion}
\label{sec:conclusion}
We have presented PLiX, an end-to-end few-shot keyword spotting system that harnesses millions of real-world audio clips to learn deep models capable of recognizing unseen spoken words (or classes) at test-time. Experimental results demonstrate that our approach achieves superior performance across $20$ languages in detecting thousands of words, showcasing its ability to generalize to novel words from as little as a single support example. The development of systems capable of seamlessly integrating human language into the fabric of emerging technologies is crucial to advancing human-computer interaction. As such, our work serves as an important step toward building practical, multilingual keyword recognizer that are highly flexible and can be personalized to user's needs. Overall, we believe few-shot meta-learning techniques can play a pivotal role in accelerating progress in the nascent area of human language understanding by machines. In the future, we plan to extend our work by integrating contextual information (such as speaker's age, gender, accent, and other environment information) to further boost accuracy, investigating privacy implications, and evaluating the approach on additional languages and dialects.

\bibliographystyle{ACM-Reference-Format}
\bibliography{main}

\end{document}